\documentclass[aps,pre,twocolumn,groupedaddress,showpacs]{revtex4-1}

\usepackage{bm}
\usepackage{graphicx}
\usepackage{color}

\begin{document}
\title{Detection of standing internal gravity waves in experiments with
convection over a wavy heated wall}
\author{L. Barel}
\author{A. Eidelman}
\author{T. Elperin}
\author{G. Fleurov}
\author{N. Kleeorin}
\author{A. Levy}
\author{I.~Rogachevskii}
\email{gary@bgu.ac.il}
\author{O.~Shildkrot}
\affiliation{\\
The Pearlstone Center for Aeronautical Engineering
Studies, Department of Mechanical Engineering,
Ben-Gurion University of the Negev, P.O.Box 653,
Beer-Sheva 84105,  Israel}
\date{\today}
\begin{abstract}
Convection over a wavy heated bottom wall in the air flow
has been studied in experiments
with the Rayleigh number $\sim 10^8$.
It is shown that the mean temperature gradient in the flow core inside
a large-scale circulation is directed upward,
that corresponds to the stably stratified flow.
In the experiments with a wavy heated bottom wall,
we detect large-scale standing internal gravity waves
excited in the regions with the stably stratified flow.
The wavelength and the period of these waves
are much larger than the turbulent spatial and time scales, respectively.
In particular, the frequencies of the observed large-scale waves vary
from 0.006 Hz to 0.07 Hz, while the turbulent time in the integral scale
is about 0.5 s.
The measured spectra of these waves contains several localized maxima,
that implies an existence of waveguide resonators 
for the large-scale standing internal gravity waves.
For comparisons, experiments with convection over a smooth plane bottom
wall at the same mean temperature difference
between bottom and upper walls have been also conducted.
In these experiments various locations with a stably stratified flow
are also found and the large-scale standing 
internal gravity waves are observed in these regions.
\end{abstract}

\maketitle

\section{Introduction}

Temperature stratified turbulence in convective and stably stratified flows
has been investigated theoretically, experimentally and in numerical simulations
due to numerous applications in geophysical, astrophysical and industrials flows.
\cite{T73,MY75,KF84,Z91,G92,N11,C14,AGL09,LX10,CS12}
One of the key ingredients of stably stratified flows are internal gravity waves.
In atmospheric and oceanic turbulence they have been a subject of intense research.
\cite{B74,GH75,M81,N02,S10,GM79,FA03,SS02,FE81,FEF84,EF84,F88,F99,EF93,C99,Z02,JSG03,BP04,BP09,JRF05,SN15}
In the atmosphere, internal gravity waves exist
at scales ranging from meters to kilometers, and are measured by direct probing
or remote sensing using radars and lidars. \cite{FA03,C99}
The sources of internal gravity waves can be flows over complex terrain, strong wind
shears, convective and other local-scale motions underlying the stably stratified
layer, and wave-wave interactions. \cite{FA03,SS02}
The internal gravity waves propagation is complicated by variable wind and density
profiles causing refraction, reflection, focusing, and ducting.

The internal gravity waves can strongly affect the small-scale turbulence.
In particular, these waves create additional productions
of turbulent energy and additional vertical turbulent fluxes of momentum and heat.
In particular, the waves emitted at a certain level, propagate
upward, and the losses of wave energy cause the production of turbulence energy.
These effects have been studied theoretically, where
the energy- and flux-budget (EFB) turbulence closure theory
which accounts for large-scale internal gravity
waves (IGW) for stably stratified atmospheric flows has been developed. \cite{ZKR09,KRZ19}
For the stationary (in statistical sense) and homogeneous turbulence,
the EFB theory without large-scale IGW yields universal
dependencies of the main turbulence parameters on the flux Richardson number
(defined as the ratio of the consumption of turbulent kinetic energy (TKE)
needed for overtaking buoyancy forces to the TKE production by the velocity shear).  \cite{ZKR07,ZKR08,ZKR10,ZKR13}
Due to the large-scale IGW, these dependencies lose their universality. \cite{ZKR09,KRZ19}
The maximum value of the flux Richardson number (universal constant
$\approx 0.25$ in the absence of the large-scale IGW)
becomes strongly variable. In the vertically homogeneous
stratification, the flux Richardson number increases with increasing wave energy
and can even exceed 1. For heterogeneous
stratification, when internal gravity waves propagate towards stronger stratification,
the maximal flux Richardson number decreases with increasing wave energy
and even can reach very small values. \cite{ZKR09,KRZ19}

Internal gravity waves also reduce the anisotropy of turbulence: in contrast
to the mean wind shear, which generates only horizontal TKE, internal gravity waves generate
both horizontal and vertical TKE.
A well-known effect of internal gravity waves
is their direct contribution to the vertical transport of momentum.
Depending on the direction of the wave propagation (downward or upward),
the internal gravity waves either strengthen or weaken the total vertical
flux of momentum. \citep{ZKR09,KRZ19}

Even in a convective turbulence,
stably stratified regions can be formed,
where the mean temperature gradient in the flow core inside
the large-scale circulation is directed upward. \cite{TSW17,NSS2000}
In these regions internal gravity waves are generated,
which affect the turbulence.
In spite of many studies of stratified turbulence and internal gravity waves,
a mechanism of formation of the stably stratified regions in convective turbulence,
and generation of internal gravity waves in these regions
is not comprehensively studied and understood.

In various flows a complex terrain (i.e., canopy and various topography)
strongly affects the stratified turbulence. \cite{B95} It changes local temperature gradients,
heat and mass fluxes, and affects a local structure of fluid flows.
Therefore, one of the important questions --- what is the effect of complex terrain
on convective and stably stratified turbulence?

To model the effect of complex terrain,
the Rayleigh-B\'{e}nard convection (RBC) with modulated boundaries has been investigated.
\cite{KP78,SWB08,WSB12,WSB14,FPZ11,TSW17,ZSVL17,RCC01}
In particular, the influence of a modulated boundary on RBC
by a lithographically fabricated periodic texture on the bottom plate
has been studied experimentally. \cite{WSB14}
The different convection patterns have been obtained by varying the Rayleigh number and the wave number of
the modulated boundary. For small Rayleigh numbers, convection takes the form of straight parallel
rolls. With increasing Rayleigh number, a secondary instability
is excited and the convection has more complex patterns. \cite{WSB14}
This secondary instability has been studied theoretically and numerically. \cite{FPZ11}

The roughness effect on the heat transport in RBC has been investigated in two-dimensional
numerical simulations by varying the height and wavelength of the roughness elements
where the sinusoidal roughness profile has been chosen. \cite{TSW17,ZSVL17}
The ultimate regime of thermal convection (when the boundary layers undergo a transition
leading to the generation of smaller scales near the boundaries
that increase the system’s efficiency in transporting the heat) has recently been found. \cite{TSW17}.
This regime in which the heat flux becomes independent of the molecular properties
of the fluid, has been predicted. \cite{K62,GL11}
The first experiment designed to use roughness to reach
the ultimate regime at accessible Rayleigh numbers has been made. \cite{RCC01}
One of the key role of the roughness elements is
the production and release of the plumes
from the roughness elements, resulting formation of
larger plumes.
This can cause an increase in the efficiency of the heat transfer. \cite{TSW17,SPV06}
The existence of two universal regimes in RBC, namely
the ultimate regime and the classical
boundary-layer-controlled regime with increased Rayleigh number,
have been demonstrated. \cite{ZSVL17}
The transition from the first to the second regime is determined
by the competition between bulk and boundary layer
flow. The bulk-dominated regime corresponds to the ultimate regime.

In the present study, we investigate another aspects related to
the roughness effect on turbulent convection. In particular, we study
formation of the stably stratified regions
and excitation of large-scale standing internal gravity waves
in laboratory experiments with
turbulent convection over a wavy heated bottom wall in the air as the working fluid.
We also compare results of these experiments to those obtained in experiments
with a smooth plane bottom wall at the same temperature difference 
between bottom and upper walls.

This paper is organized as follows.
In Section II we present theoretical analysis which allows us to determine
the frequencies of the standing internal gravity waves
in stably stratified flow.
In Section III we describe the experimental set-up and instrumentation
for laboratory study of the internal gravity waves.
The results of the laboratory experiments are
discussed in Section IV. Finally, conclusions are drawn in Section V.

\section{Internal gravity waves}

Let us first consider the internal gravity waves in a stably stratified fluid flow
in the absence of turbulence and neglecting dissipations.
These waves are described by the linearized momentum
and entropy equations written in the Boussinesq approximation \cite{T73,M81,N02,S10}
\begin{eqnarray}
{\partial {\bm V}^{\rm W} \over \partial t} &=& - {{\bm \nabla} P^{\rm W} \over \rho_{\rm eq}} + g \, S^{\rm W} {\bm e} ,
 \label{B1}\\
{\partial S^{\rm W} \over \partial t} &=&  - g^{-1} N^2 \, {\bm V}^{\rm W} \cdot {\bm e},
 \label{B2}
\end{eqnarray}
where
\begin{eqnarray}
S^{\rm W}=c_{\rm v} \left[(1-\gamma) \, {P^{\rm W} \over P_{\rm eq}} + \gamma \, {T^{\rm W} \over T_{\rm eq}} \right],
 \label{SB1}
\end{eqnarray}
the wave fields ${\bm V}^{\rm W}$, $P^{\rm W}$, $S^{\rm W}$ and $T^{\rm W}$
are the perturbations of the velocity, pressure, entropy and temperature,
respectively,  ${\bm e}$ is the vertical unit vector, ${\bm g}$ is the acceleration due to gravity,
and $\gamma=c_{\rm p}/c_{\rm v}$ is the ratio of the specific heats, $c_{\rm p}$ and $c_{\rm v}$ are the specific heats at constant pressure and volume, respectively. The Brunt-V\"{a}is\"{a}l\"{a} frequency is $N(z)= (g \nabla_z S_{\rm eq})^{1/2}$, where
\begin{eqnarray}
\nabla_z S_{\rm eq}=c_{\rm v} \left[(1-\gamma) \, \nabla_z \ln P_{\rm eq} + \gamma \, \nabla_z \ln T_{\rm eq} \right],
 \label{SB2}
\end{eqnarray}
and the fields $S_{\rm eq}$, $P_{\rm eq}$, $T_{\rm eq}$ and $\rho_{\rm eq}$ are the entropy, pressure temperature and density at an equilibrium given by the following equations: ${\bm V}_{\rm eq}=0$
and ${\bm \nabla} P_{\rm eq} = \rho_{\rm eq} \, {\bm g}$, where ${\bm V}_{\rm eq}$ is the velocity at an equilibrium.
For conditions pertinent to the laboratory experiments discussed in the next sections,
${\bm \nabla} P_{\rm eq} \approx 0$, so that
$\nabla_z S_{\rm eq}\approx c_{\rm p} \, \nabla_z \ln T_{\rm eq}$.
The classical Boussinesq approximation with div ${\bm V}^{\rm W} =0$ is applied here.

Equations~(\ref{B1}) and~(\ref{B2}) yield the frequency $\omega$ of the internal gravity waves:
\begin{eqnarray}
\omega = N(z) {k_h \over k} ,
\label{B11}
\end{eqnarray}
where ${\bm k}={\bm k}_h + {\bm e} k_z$  is the wave vector and ${\bm k}_h=(k_x, k_y)$  is the wave vector in the horizontal direction.
Propagation of the internal gravity waves in the stably stratified flow in the approximation of geometrical optics is determined by the following Hamiltonian equations:
\begin{eqnarray}
{\partial {\bm r} \over \partial t} &=& {\partial \omega \over \partial {\bm k}},
\label{B16}\\
{\partial {\bm k} \over \partial t} &=& - {\partial \omega \over \partial {\bm r}},
\label{B10}
\end{eqnarray}
where  ${\bm r}$ is the radius-vector of the centre of the wave packet. \cite{W62})  Since the Brunt-V\"{a}is\"{a}l\"{a} frequency $N=N(z)$, the only non-zero spatial derivative, $\nabla_z \omega \not=~0$, is in the vertical direction. Therefore, Eq.~(\ref{B10}) yields ${\bm k}_h=$ const. The vertical component of the wave vector $k_z=k_z(z)$ is determined from Eq.~(\ref{B11}):
\begin{eqnarray}
k_z(z)=k_h \left({N^2(z) \over \omega^2} -1\right)^{1/2} .
\label{B12}
\end{eqnarray}

We perform standard calculations by taking twice {\bf curl}
to exclude the pressure term in Eq.~(\ref{B1}), calculating the
time derivative of the obtained equation and using Eq.~(\ref{B2}).
This procedure yields the following equation:
\begin{eqnarray}
{\partial^2 \over \partial t^2} \, \Delta {\bm V}^{\rm W}  &=& \left[{\bm \nabla} ({\bm e} \cdot {\bm \nabla}) - {\bm e} \Delta \right] V^{\rm W}_z N^2(z) ,
\label{B3}
\end{eqnarray}
which is equivalent to the system of equations for the vertical and horizontal velocity components:
\begin{eqnarray}
{\partial^2 \over \partial t^2} \, \Delta V_z^{\rm W} &=& - N^2(z) \Delta_\perp V^{\rm W}_z ,
 \label{B4}\\
 {\partial^2 \over \partial t^2} \, \Delta {\bm V}^{\rm W}_\perp  &=& {\bm \nabla}_\perp \nabla_z^2 \left[V^{\rm W}_z N^2(z) \right] .
 \label{B5}
\end{eqnarray}
Here ${\bm V}^{\rm W}={\bm V}^{\rm W}_\perp + {\bm e} V_z^{\rm W}$ with ${\bm V}^{\rm W}_\perp=(V^{\rm W}_x, V^{\rm W}_y)$ being the horizontal velocity and $\Delta=\Delta_\perp + \nabla_z^2$.
Solution of Eq.~(\ref{B4}) for the vertical velocity $V_z^{\rm W}(t,{\bm r})$
we seek for in the form of standing internal gravity waves existing
in the range $z_{\rm min} \leq z \leq z_{\rm max}$:
\begin{eqnarray}
&&V_z^{\rm W}(t,{\bm r}) = V_\ast \cos (\omega t) \cos ({\bm k}_h \cdot {\bm r})
\nonumber\\
&& \quad \quad \times \sin \left(\int_{z_{\rm min}}^z k_z(z') \,dz' + \varphi\right) ,
\label{B6}
\end{eqnarray}
where $V_\ast$ is the amplitude of the vertical velocity for the wave field.
Substituting Eq.~(\ref{B6}) into Eq.~(\ref{B5}), we determine perturbations of
the horizontal velocity ${\bm V}^{\rm W}_\perp(t,{\bm r})$ for the wave field as
\begin{eqnarray}
&& {\bm V}^{\rm W}_\perp(t,{\bm r}) = - {\bm k}_h {k_z(z) \over k_h^2} \,
V_\ast \cos (\omega t) \sin ({\bm k}_h \cdot {\bm r})
\nonumber\\
&& \quad \quad \times \cos \left(\int_{z_{\rm min}}^z k_z(z') \,dz'+ \varphi\right) .
\label{B8}
\end{eqnarray}
Using Eqs.~(\ref{B2}) and~(\ref{B6}),
we obtain solution for perturbations of the entropy $S^{\rm W}(t,{\bm r})$:
\begin{eqnarray}
&& S^{\rm W}(t,{\bm r}) =  {N^2(z) \over g \, \omega} \,
V_\ast \sin (\omega t) \cos ({\bm k}_h \cdot {\bm r})
\nonumber\\
&& \quad \quad \times \sin \left(\int_{z_{\rm min}}^z k_z(z') \,dz'+ \varphi\right) .
\label{B9}
\end{eqnarray}

The solution~(\ref{B6}) should satisfy the following boundary conditions: $V_z^{\rm W}(z=z_{\rm min})=0$
and $V_z^{\rm W}(z \approx z_{\rm max})=0$.
The latter boundary condition at the vicinity of $z \approx z_{\rm max}$ implies that
\begin{eqnarray}
\int_{z_{\rm min}}^{z_{\rm max}} k_z(z') \,dz' = \pi \left(m + {1 \over 4} \right) ,
\label{B7}
\end{eqnarray}
where $z_{\rm max}$ is the reflection (or ''turning") point in which $k_z(z=z_{\rm max})=0$.
To get the condition~(\ref{B7}), we use an analogy with the behavior of the wave function
near the turning points in the semi-classical limit applied in quantum mechanics.
\cite{LL13,DA65} The wave functions are described in terms of the Airy functions, and
Eq.~(\ref{B7}) is analogous to the Bohr-Sommerfeld quantization condition. In particular,
the phase $\varphi$ in Eqs.~(\ref{B6})--(\ref{B9}) is determined using asymptotic
solution of Eq.~(\ref{B4}) in the vicinity of the turning points:
\begin{eqnarray}
V_z^{\rm W} &=& {1 \over 2} \, \xi^{-1/4} \, \exp \left(- {2 \over 3} \xi^{3/2}\right) , \; \;  {\rm for} \; \;
\xi \gg 1,
\nonumber\\
V_z^{\rm W} &=& {1 \over 2} \, |\xi|^{-1/4} \, \sin \left({2 \over 3} |\xi|^{3/2} + {\pi \over 4} \right) , \; \; {\rm for} \; \; |\xi| \ll 1,
\nonumber\\
\label{B20}
\end{eqnarray}
where $\xi \propto z_{\rm max} - z$.
Equation~(\ref{B7}) allows us to determine the frequencies of the standing internal
gravity waves for the quadratic profile of the Brunt-V\"{a}is\"{a}l\"{a}
frequency $N^2(z) = N^2_0(1-z^2/L_N^2)$. In particular, calculating the integral
$\int_{z_{\rm min}}^{z_{\rm max}} k_z(z') \,dz'$, we obtain:
\begin{eqnarray}
\omega_m &=& {N_0 \over k_h L_N} \biggl\{\biggl[\left(m + {1 \over 4}\right)^2 + \left(k_h L_N\right)^2 \biggr]^{1/2}
\nonumber\\
&&- \left(m + {1 \over 4}\right)\biggr\} ,
\label{B14}
\end{eqnarray}
where $L_N$ is the characteristic scale of the Brunt-V\"{a}is\"{a}l\"{a} frequency variations,
and $m=0, 1, 2, ...$. Equation~(\ref{B14}) implies the existence of a discrete spectrum
of the standing internal gravity waves.
In the long wavelength limit, $k_h L_N \ll 1$, Eq.~(\ref{B14}) yields
\begin{eqnarray}
\omega_m = {2 N_0 \over 4 m +1} k_h L_N .
\label{B15}
\end{eqnarray}
We will apply Eq.~(\ref{B14}) in the experimental study of convection over a wavy
heated bottom surface, where the large-scale internal gravity waves are excited
in the stably stratified regions formed inside the flow core of
the large-scale circulation.

\section{Experimental set-up}

In this section we describe the experimental
set-up.
The experiments have been conducted in air as the working fluid
in rectangular chamber with dimensions $L_x \times L_y \times L_z$,
where $L_x = L_z = 26$ cm, $L_y=56$ cm and the axis $z$ is in the vertical
direction.
The side walls of the chamber are made
of transparent Perspex with the thickness of $1$
cm.

A vertical mean temperature gradient in the turbulent air flow is formed by attaching two aluminium heat exchangers to the bottom and top walls of the test section (a heated bottom and a cooled top wall of the chamber). A thickness of the massive aluminium heat exchangers is 2 cm. The top plate is a bottom wall of the tank with cooling water. Cold water is pumped into the cooling system through two inlets and flows out through two outlets located at the side wall of the cooling system.
The bottom plate is attached to the electrical heater with wire tightly laid in the grooves milled in the aluminum plate and provided uniform heating. Energy supplied to the heater is varied in order to obtain necessary temperature difference between heater and cooler. Characteristic time of heater is approximately 90 min that stabilize applied temperature during measurements.

\begin{figure}
\centering
\includegraphics[width=7.0cm]{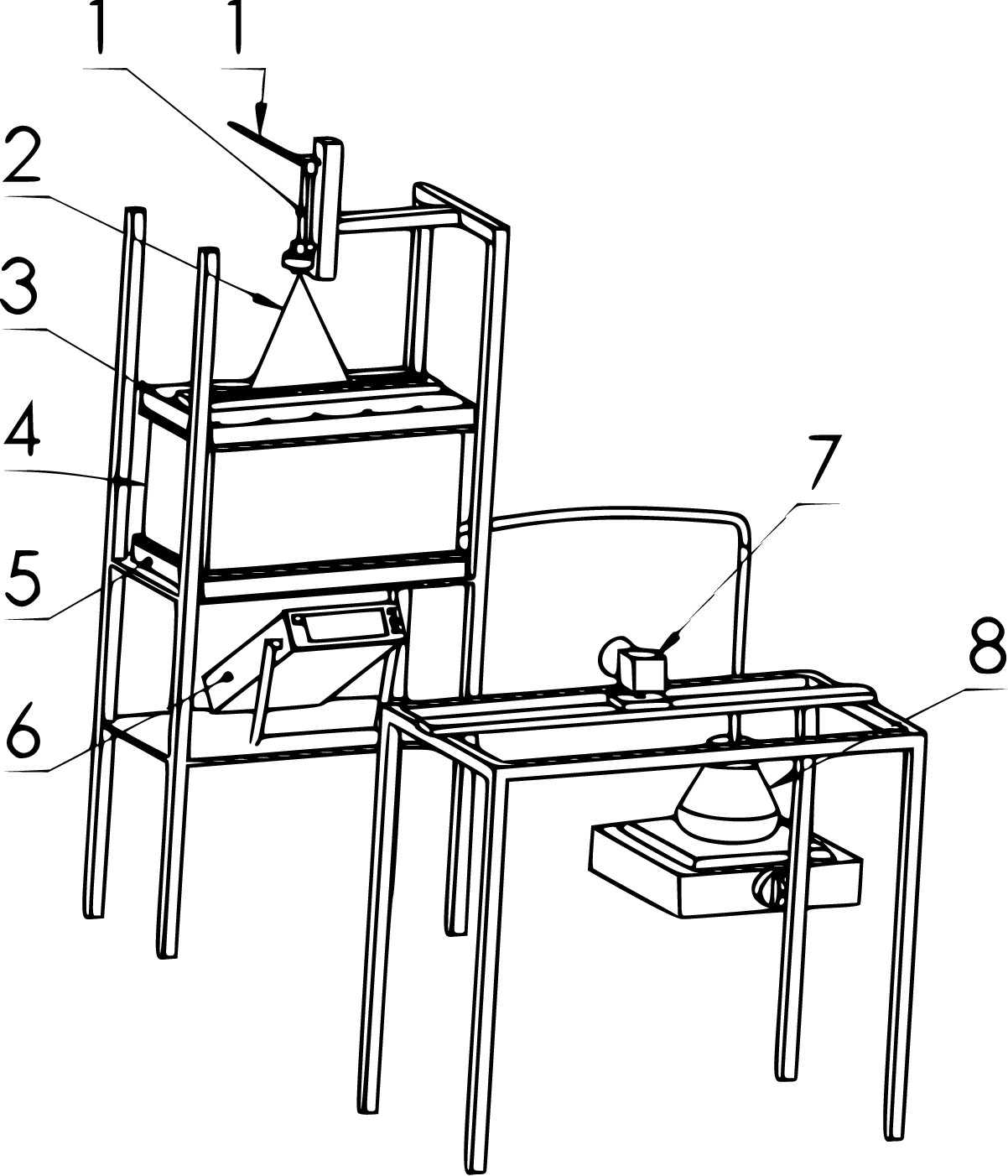}
\caption{\label{Fig1}
A sketch of the experimental set-up:
the optics (1) and the laser light sheet (2) of the PIV system;
the chamber (4); the cooler (3) at the top surface
and the heater (5) at the bottom surface;
the data recorder (6) for temperature measurement system; the CCD camera (7)
and the generator of incense smoke (8) for the PIV measurements.
}
\end{figure}

\begin{figure}
\centering
\includegraphics[width=8.0cm]{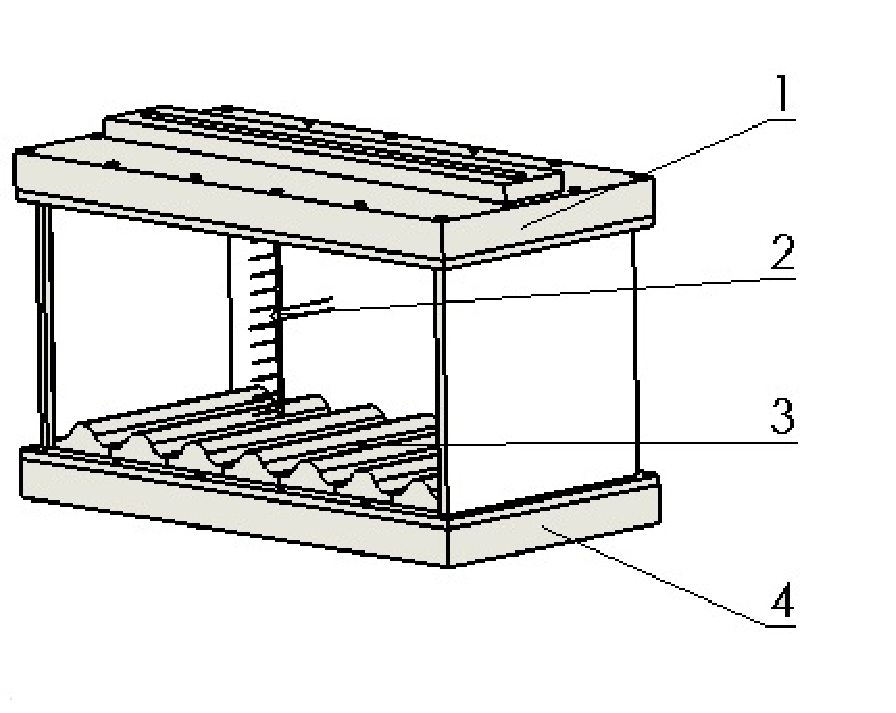}
\caption{\label{Fig2}
A sketch of the chamber: the cooler (1) at the top surface;
the heater (4) at the bottom surface; the temperature probe (2) equipped with 12
E-thermocouples; a wavy bottom surface (3) with a sinusoidal modulation.
}
\end{figure}

We study the effects of complex terrain
on the structure of the velocity and temperature
fields in temperature stratified turbulence.
In the laboratory experiments complex terrain is modelled
by a wavy bottom surface of the chamber
which is manufactured from aluminium.
It is produced from a plate of a thickness 4 cm.
The wavy bottom surface has a sinusoidal modulation containing
7 periods with wavelength 8 cm and amplitude 1 cm.
A sketch of the experimental set-up is shown in Figs.~\ref{Fig1} and~\ref{Fig2}.
The results of these experiments are compared to those obtained
in experiments with a smooth plane bottom surface at the same temperature difference
between the top and bottom plates.

The temperature field is measured with  a
temperature probe equipped with 12
E-thermocouples (with the diameter of 0.13 mm and
the sensitivity of $\approx 65 \, \mu$V/K)
attached to a rod with a diameter 4 mm.
The spacing between thermocouples along the rod
is 22 mm. Each thermocouple is inserted into a
1 mm diameter and 45 mm long case. A tip of a
thermocouple protruded at the length of 15 mm out
of the case.
Thermocouples of type {\it E} are used for the temperature measurements
in the core flow, while thermocouples of type {\it K} are used
for temperature measurements at the heater and the cooler.
All thermocouples are built by a manufacturer (''The temperature handbook",
Omega Engineering inc., Stanford, 2011).
Calibrations of all E-thermocouples in the temperature probe
have been performed for three experiments with boiled water
$(T=373$ K), the cold water with ice $(T=273$ K) and
water with intermediate temperature $(T=296$ K).
Comparisons have been performed using a precision temperature
measurement manufactured device.

The temperature is measured for 11
rod positions with 20 mm intervals in the
horizontal direction.
This probe is used in the experiments with
smooth plane bottom surface, while a temperature probe equipped with 13
E-thermocouples is used in the experiments with
the wavy bottom surface.
A sequence of 500 temperature readings for every
thermocouple at every rod position is recorded
and processed.
We have measured the temperature field in many locations.
Performing direct continuous measurements of the temperatures
at the cooled top surface and at the
heated bottom surface and using a standard device for supporting constant
temperature difference $\Delta T$ between the top and bottom surfaces (Contact voltage regulator TDGC-2K),
we control the constant temperature difference $\Delta T$ during the experiments.

The velocity field is measured using a
Stereoscopic Particle Image  Velocimetry.
\cite{AD91,RWK07,W00} In the
experiments we use LaVision Flow Master III
system. A double-pulsed light sheet is provided
by a Nd-YAG laser (Continuum Surelite $ 2 \times
170$ mJ). The light sheet optics includes
spherical and cylindrical Galilei telescopes with
tuneable divergence and adjustable focus length.
We use the progressive-scan 12 bit digital CCD
camera (with pixel size $6.7 \, \mu$m $\times \,
6.7 \, \mu$m and $1280 \times 1024$ pixels) with
a dual-frame-technique for cross-correlation
processing of captured images. A programmable
Timing Unit (PC interface card) generated
sequences of pulses to control the laser, camera
and data acquisition rate.

An incense smoke with sub-micron particles  is used as a
tracer for the PIV measurements. Smoke is
produced by high temperature sublimation of solid
incense grains. Analysis of smoke particles using
a microscope (Nikon, Epiphot with an
amplification of 560) and a PM-300 portable laser
particulate analyzer shows that these particles
have an approximately spherical shape and that
their mean diameter is of the order of $0.7
\mu$m. The maximum tracer particle displacement in
the experiment is of the order of $1/4$ of the
interrogation window. The average displacement of
tracer particles is of the order of $2.5$
pixels. The average accuracy of the velocity
measurements is of the order of $4 \%$ for the
accuracy of the correlation peak detection in the
interrogation window of the order of $0.1$ pixel.
\cite{AD91,RWK07,W00}

We determine the mean and the
r.m.s.~velocities, two-point correlation
functions and an integral scale of turbulence
from the measured velocity fields. Series of 520
pairs of images acquired with a frequency of 2
Hz, are stored for calculating velocity maps and
for ensemble and spatial averaging of turbulence
characteristics. The center of the measurement
region coincides with the center of
the chamber. We measure velocity in a flow
domain $256 \times 503$ mm$^2$ with a spatial
resolution of 393 $\mu$m / pixel. The velocity field in the probed region
is analyzed with interrogation windows of $32
\times 32$ pixels. In every
interrogation window a velocity vector is
determined from which velocity maps comprising
$27 \times 53$ vectors are
constructed.
The mean and r.m.s. velocities for
every point of a velocity map are calculated by
averaging over 520 independent velocity maps, and
the obtained averaged velocity map is
averaged also over the central flow region.

The two-point correlation functions  of the
velocity field are determined for every point of
the central part of the velocity map (with $16
\times 16$ vectors) by averaging over 520
independent velocity maps, which yields 16
correlation functions in horizontal and vertical directions.
The two-point correlation function is
obtained by averaging over the ensemble of these
correlation functions.
An integral scale of turbulence, $\ell$, is determined from the
two-point correlation functions of the velocity
field. The turbulence time scale at the
integral scale is $\tau = \ell / \sqrt{\langle {\bf u}^2 \rangle}$,
where ${\bf u}$ are the velocity fluctuations and
$\sqrt{\langle {\bf u}^2 \rangle}$ is the r.m.s. of the velocity fluctuations.
In the experiments we evaluated the
variability between the first and the last 20
velocity maps of the series of the measured
velocity field. Since very small variability is
found, these tests show that 520 velocity maps
contain enough data to obtain reliable
statistical estimates.
The size of the probed region does not affect our results.

In the experimental study we employ a triple decomposition
whereby the instantaneous temperature $T^{\rm tot}=T
+ \theta$, where $\theta$ are the temperature
fluctuations and $T$ is the temperature
determined by sliding averaging of the
instantaneous temperature field over the time (5~s),
that is by one order of magnitude larger than the
characteristic turbulence time (0.5~s). The
temperature $T$ is given by a sum, $T
=\overline{T} + T^{\rm W}$, where $T^{\rm W}$ are
the long-term variations of the temperature $T$
due to the large-scale standing internal gravity waves
around the mean value $\overline{T}$. The mean
temperature $\overline{T}$ is obtained by the
additional averaging of the temperature $T$ over
the time 500~s.

The time interval during which
temperature field is measured at every point, is $500$ s.
In the temperature measurements, the acquisition
frequency of the temperature is $1.25$ Hz, and the
corresponding acquisition time is $0.8$ s.
It is larger than the characteristic turbulence
time (see below), and is much smaller than the
period of the long-term oscillations of the mean
temperature caused by internal gravity waves.
Therefore, the acquisition frequency of temperature is high
enough to provide sufficiently long time series
for statistical estimation of the mean
temperature $\overline{T}$ and the long-term
variations $T^{\rm W}$ of the temperature
due to the large-scale internal gravity waves.
Similar experimental set-ups, measurement techniques
for temperature and velocity fields
and data processing procedures have been used
previously in our experimental study of different
aspects of turbulent convection
\cite{EEKR06,BEKR09} and stably stratified turbulence. \cite{BEKR11,EEKR13}

\section{Experimental results}

In this section we describe experimental results
related to formation of stably stratified regions
in turbulent convection
and excitation of internal gravity waves.
We perform two sets of experiments with
a wavy bottom surface and a smooth plane bottom
surface for the same imposed mean temperature difference $\Delta T$
between the bottom and upper surfaces of the chamber.
The velocity measurements show that in the both
sets of experiments, a single large-scale circulation
is observed for the temperature differences $\Delta T= 40$~K and
$\Delta T= 50$ K.

\begin{figure}
\centering
\includegraphics[width=7.5cm]{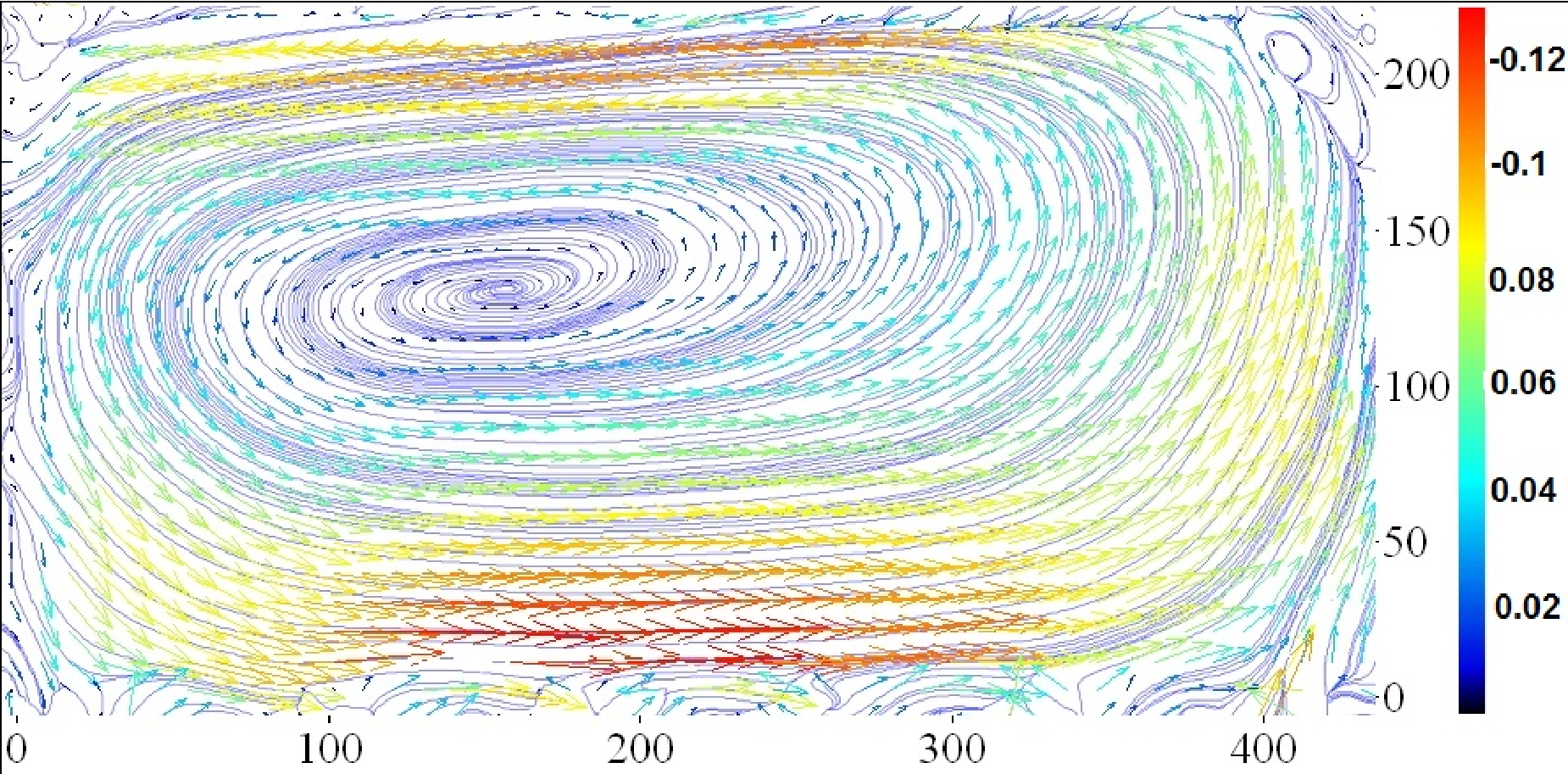}
\includegraphics[width=7.5cm]{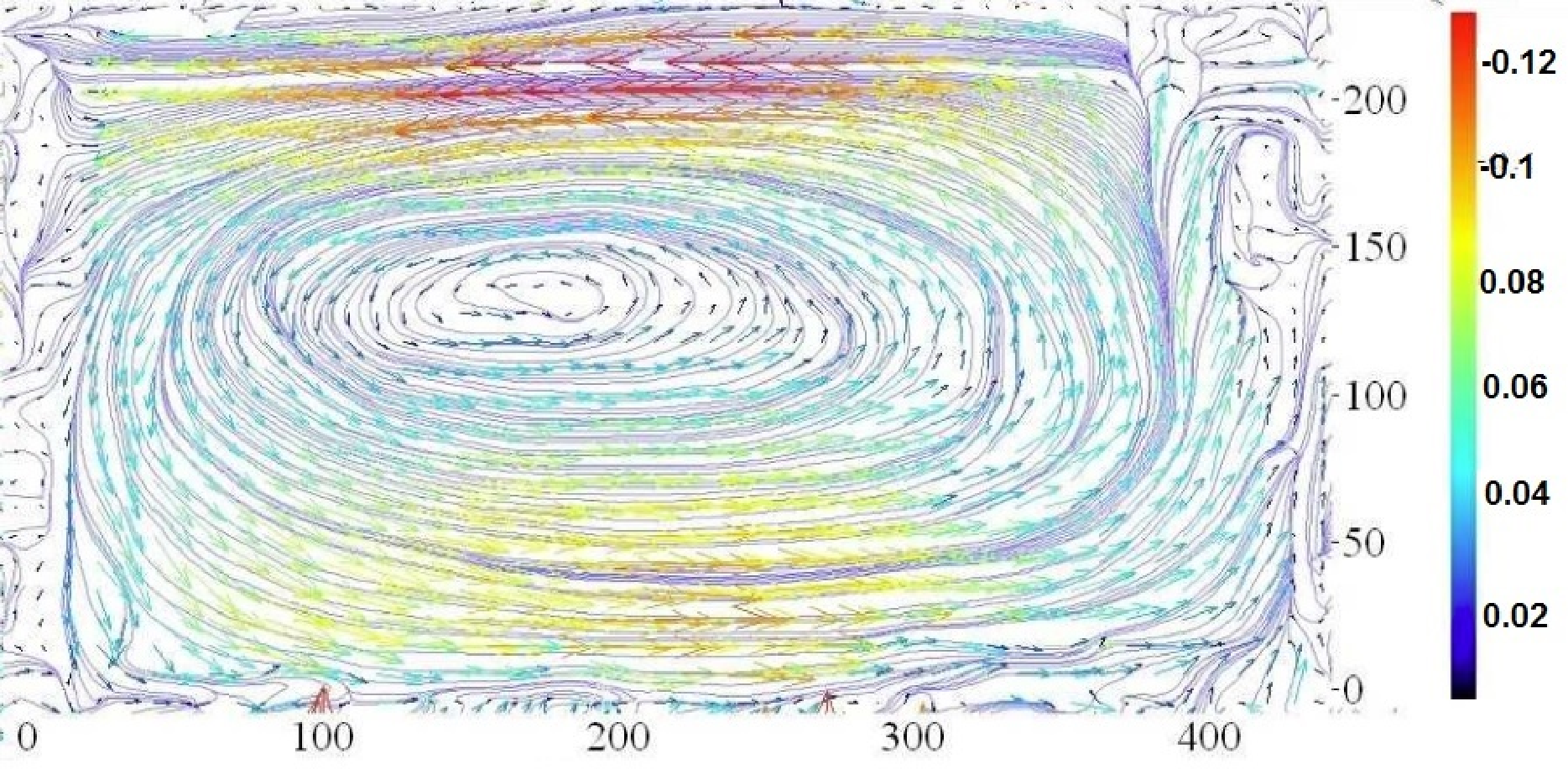}
\caption{\label{Fig3}
The mean velocity patterns in the $yz$ plane for the experiments with the wavy bottom surface
(upper panel) and the smooth plane bottom surface (bottom panel) obtained at $\Delta T= 50$ K.
Here $z$ and $y$ are measured in mm and velocity is measured in m/s.
}
\end{figure}

In Fig.~\ref{Fig3} we show the mean velocity patterns obtained
in the experiments with the wavy bottom surface
(upper panel) and smooth plane bottom surface (bottom panel).
A difference in the mean velocity patterns is seen at the vicinity
of the wavy bottom surface where the flows with the sinusoidal modulation
of the mean velocity field are observed
(see upper panel in Fig.~\ref{Fig3}).
For lower and higher temperature differences
between the bottom and top surfaces of the chamber,
two large-scale circulations are observed and
the mean temperature field is more complicated.
This case will be investigated in a separate study.
Experiments with smaller $L_z$ increases aspect ratio of the chamber,
where more than two large-scale circulations can be observed
and the mean temperature field will be much more complicated.
This case will be also investigated in a separate study.

Let us discuss parameters in the experiments.
The characteristic turbulence time is $\tau = 0.28 - 0.62$ seconds, while
the characteristic period for the
large-scale circulatory flow is about $10$ seconds, which is by
order of magnitude larger than the turbulence time $\tau$.
These two characteristic times are much smaller
than the time during which the velocity fields
are measured $(\sim 260$ seconds).
The maximum Rayleigh number, ${\rm Ra}=
\alpha \, g \, L_z^3 \, \Delta T /(\nu \, \kappa)$,
in the turbulent convection is about $10^8$, where $\alpha$ is the thermal
expansion coefficient, $\nu$ is the kinematic viscosity,
$\kappa$ is the thermal diffusivity,
and $L_z$ is the height
measured from the lower point of a wavy surface
with the sinusoidal modulation to the upper surface of the chamber.

\begin{figure}
\centering
\includegraphics[width=9.0cm]{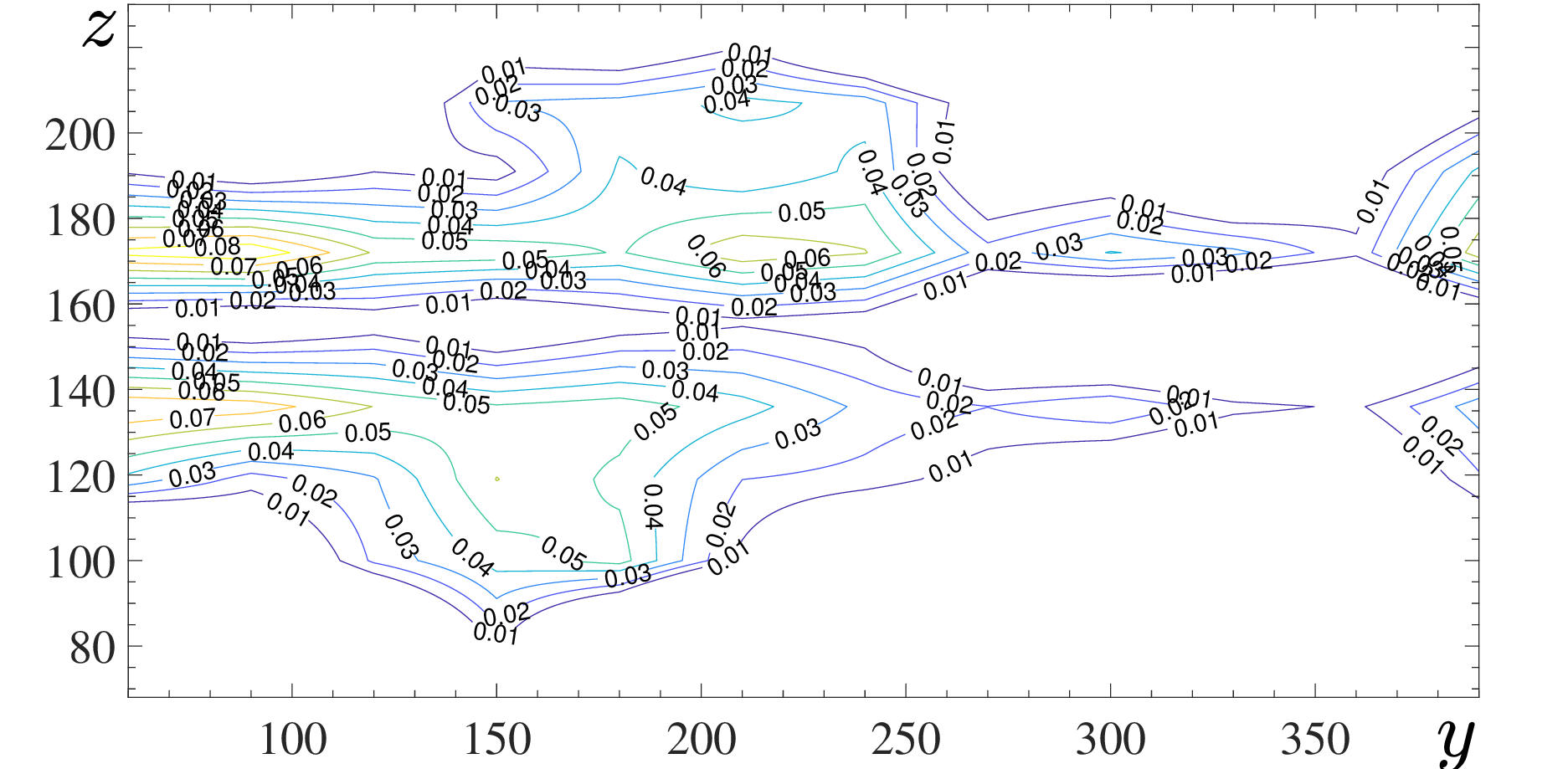}
\caption{\label{Fig4}
Isolines in $yz$ plane of the vertical mean temperature gradient $\nabla_z \overline{T}$
inside the large-scale circulation in the experiments with
the wavy bottom surface for $\Delta T = 50$ K.
Only the regions with positive vertical mean temperature gradient are shown.
Here $y$ and $z$ are measured in mm, and
the mean temperature gradient is measured in K/cm.
}
\end{figure}

\begin{figure}
\centering
\includegraphics[width=8.7cm]{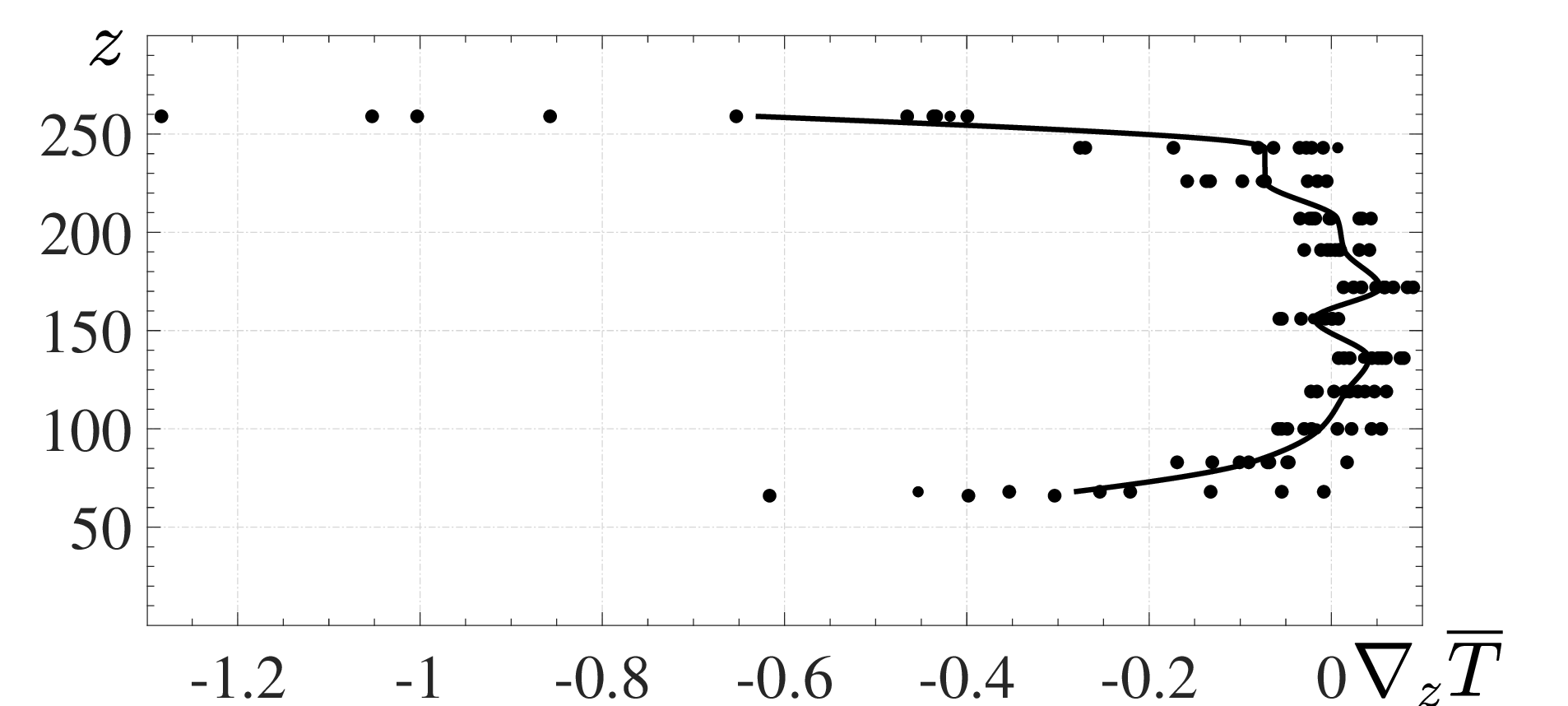}
\caption{\label{Fig5}
Mean profile of the vertical mean temperature gradient $\nabla_z \overline{T}$
in the experiments with the wavy bottom surface for $\Delta T = 50$ K
obtained by averaging over 15 vertical profiles measured in different cross-sections
for different $y$
inside the large-scale circulation.
Here the height $z$ is measured in mm, and
the mean temperature gradient is measured in K/cm.
}
\end{figure}

\begin{figure}
\centering
\includegraphics[width=9.0cm]{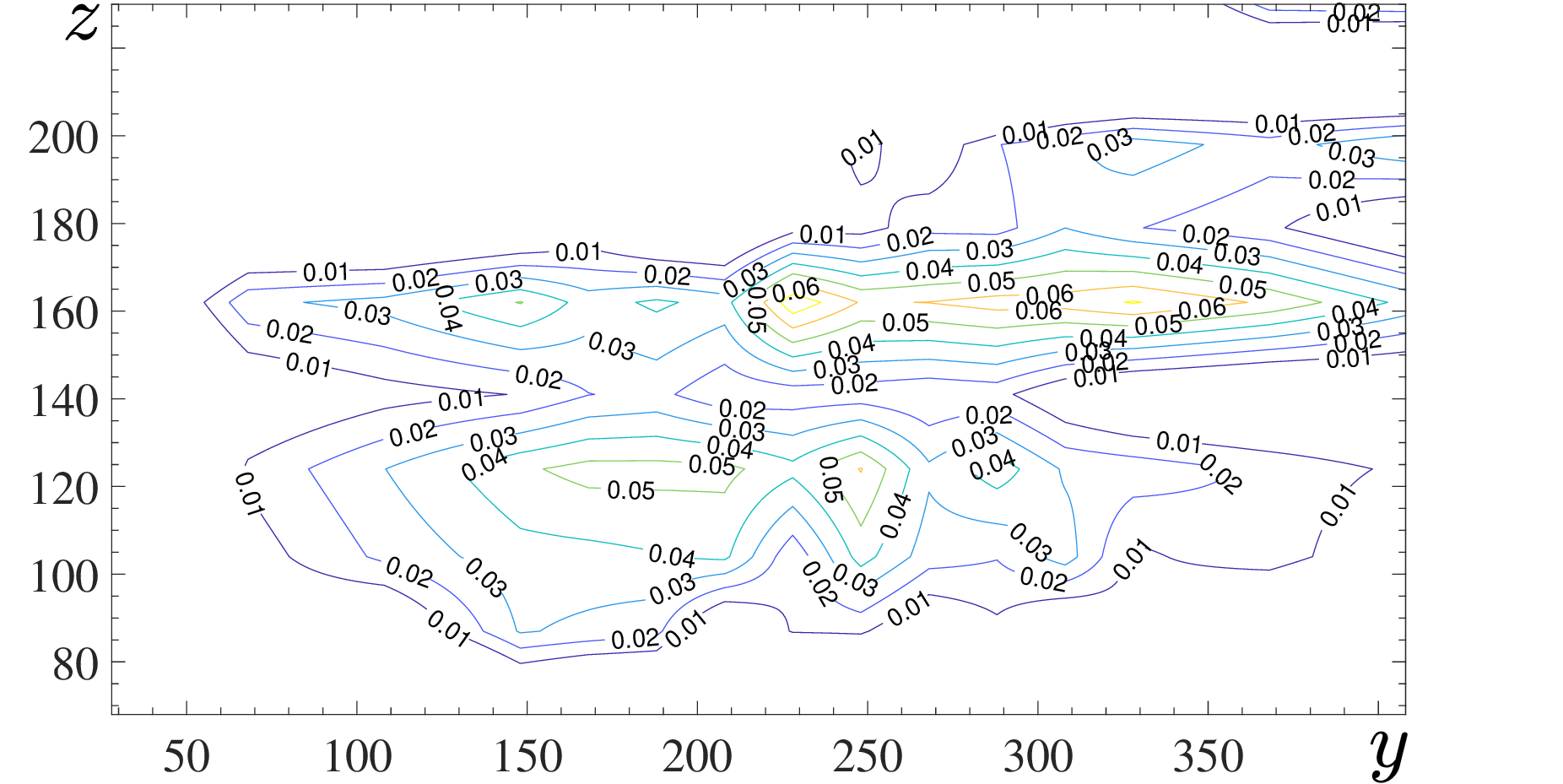}
\caption{\label{Fig6}
Isolines in $yz$ plane of the vertical mean temperature gradient $\nabla_z \overline{T}$
inside the large-scale circulation in the experiments with
the wavy bottom surface for $\Delta T = 40$ K.
Only the regions with positive vertical mean temperature gradient are shown.
Here $y$ and $z$ are measured in mm, and
the mean temperature gradient is measured in K/cm.
}
\end{figure}

\begin{figure}
\centering
\includegraphics[width=8.7cm]{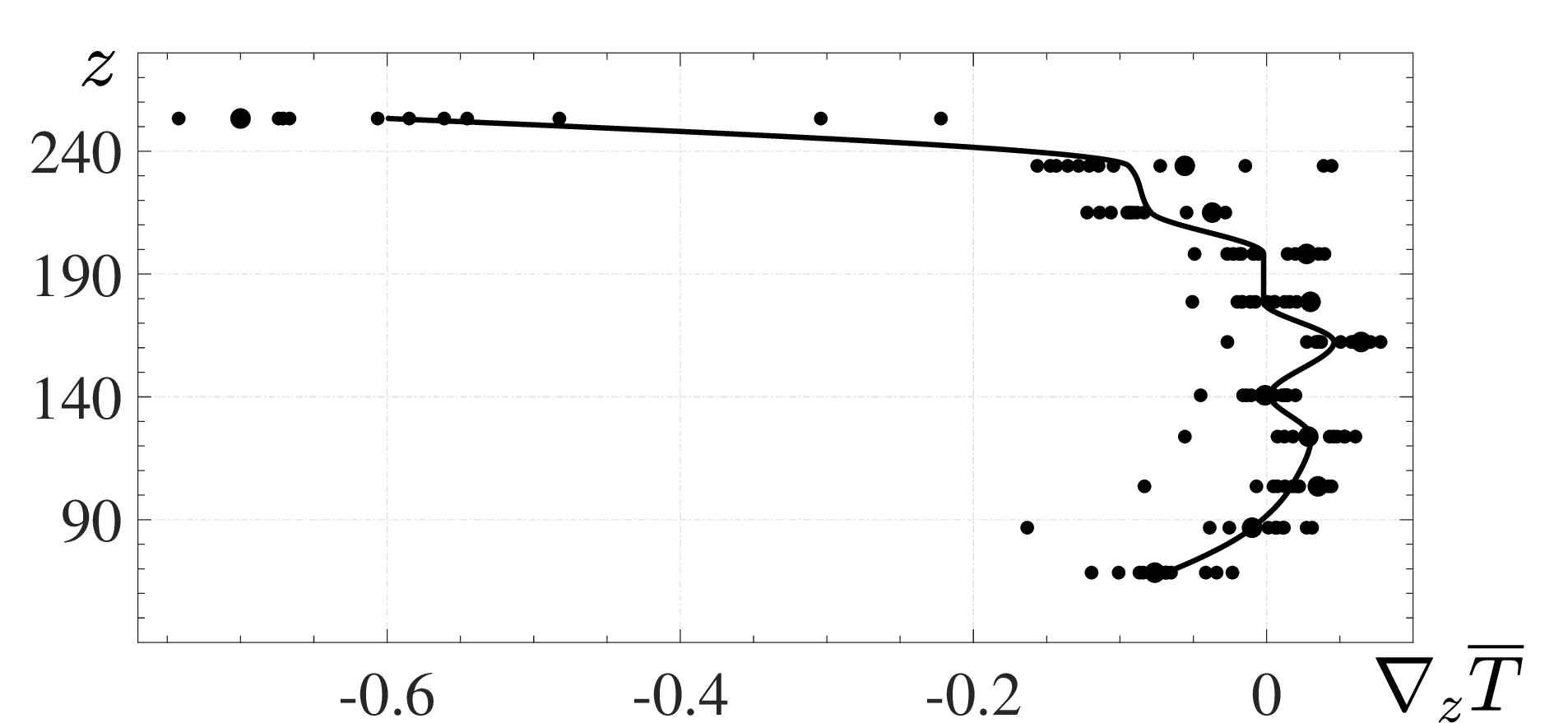}
\caption{\label{Fig7}
Mean profile of the vertical mean temperature gradient $\nabla_z \overline{T}$
in the experiments with the wavy bottom surface for $\Delta T = 40$ K
obtained by averaging over 15 vertical profiles measured in different cross-sections
for different $y$  inside the large-scale circulation.
Here the height $z$ is measured in mm, and
the mean temperature gradient is measured in K/cm.
}
\end{figure}

The temperature measurements in the experiments with
a wavy bottom surface show that the mean temperature gradient
in the flow core inside the large-scale circulation is directed upward
($\nabla_z \overline{T} > 0$),
that corresponds to a stably stratified flow.
For instance, in Figs.~\ref{Fig4} and~\ref{Fig6} we plot
isolines of the vertical mean temperature gradient $\nabla_z \overline{T}$
inside the large-scale circulation in $yz$ plane
where we show the regions with the positive
vertical mean temperature gradient only.
In Figs.~\ref{Fig5} and~\ref{Fig7} we also plot mean profiles of the
vertical mean temperature gradient $\nabla_z \overline{T}$
obtained in these experiments by averaging over 15 vertical profiles
measured in different cross-sections
for different $y$ inside the large-scale circulation.
We discuss here the results of the experiments performed for two values of the
temperature difference $\Delta T$ between the bottom and top surfaces:
$\Delta T = 50$~K (Figs.~\ref{Fig4} and~\ref{Fig6})
and $\Delta T = 40$ K (Figs.~\ref{Fig5} and~\ref{Fig7}).
All measurements have been done in the plane $x=0.5 L_x$.
The accuracy of the mean temperature measurements is about 0.07 \%
(i.e. the accuracy is about 0.2 K for typical mean temperatures $\approx 300$ K).
The accuracy of the mean temperature gradient measurements is about 5 \%
for $\nabla_z \overline{T} \approx 0.1$ K/cm,
and it is about 10 \% for $\nabla_z \overline{T} \approx 0.01$ K/cm.

As can be seen in Fig.~\ref{Fig4}, there are two regions with positive
mean temperature gradient separated by the region with negative mean temperature gradient
in the center of the core flow.
In the regions with negative vertical mean temperature gradient,
the value of $N^2 \leq 0$ and the internal gravity waves cannot exist in these regions.
This means that in the experiments with $\Delta T= 50$ K, there are two waveguides for the internal gravity waves. On the other hand, as can be seen in Fig.~\ref{Fig6} ($\Delta T= 40$~K) there is one large region with positive
mean temperature gradient which corresponds to a one waveguide.
Similar trends can be seen in Figs.~\ref{Fig5} and~\ref{Fig7}.
In these experiments the maximum gradient Richardson number, Ri $=N^2/{\rm Sh}^2$, based
on the mean velocity shear, ${\rm Sh}$, of the large-scale circulation is about 1.

\begin{figure}
\centering
\includegraphics[width=9.0cm]{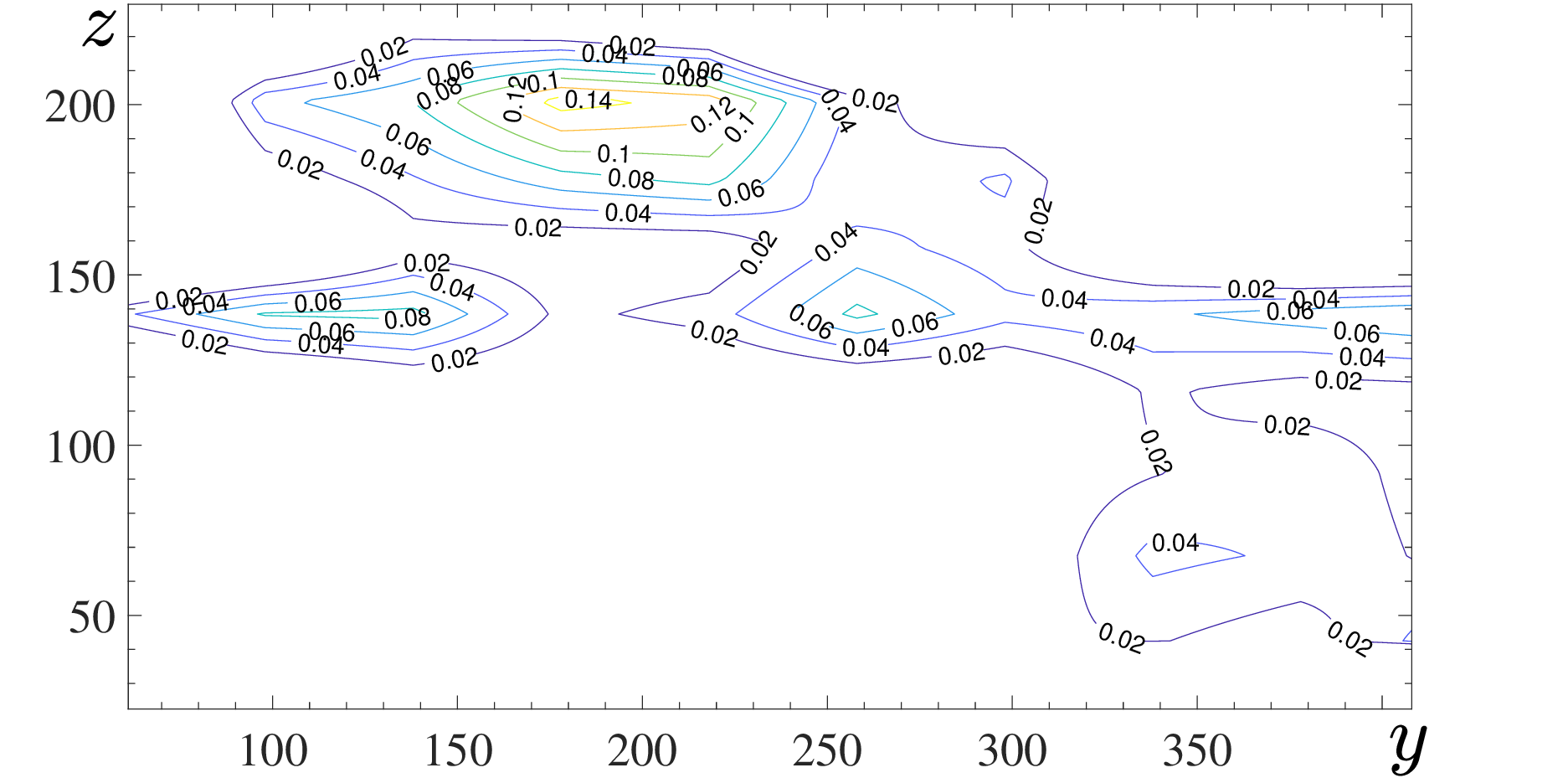}
\caption{\label{Fig8}
Isolines in $yz$ plane of the vertical mean temperature gradient $\nabla_z \overline{T}$
inside the large-scale circulation in the experiments
with the smooth plane bottom surface for $\Delta T = 50$ K.
Only the regions with positive vertical mean temperature gradient are shown.
Here $y$ and $z$ are measured in mm, and
the mean temperature gradient is measured in K/cm.
}
\end{figure}

\begin{figure}
\centering
\includegraphics[width=8.7cm]{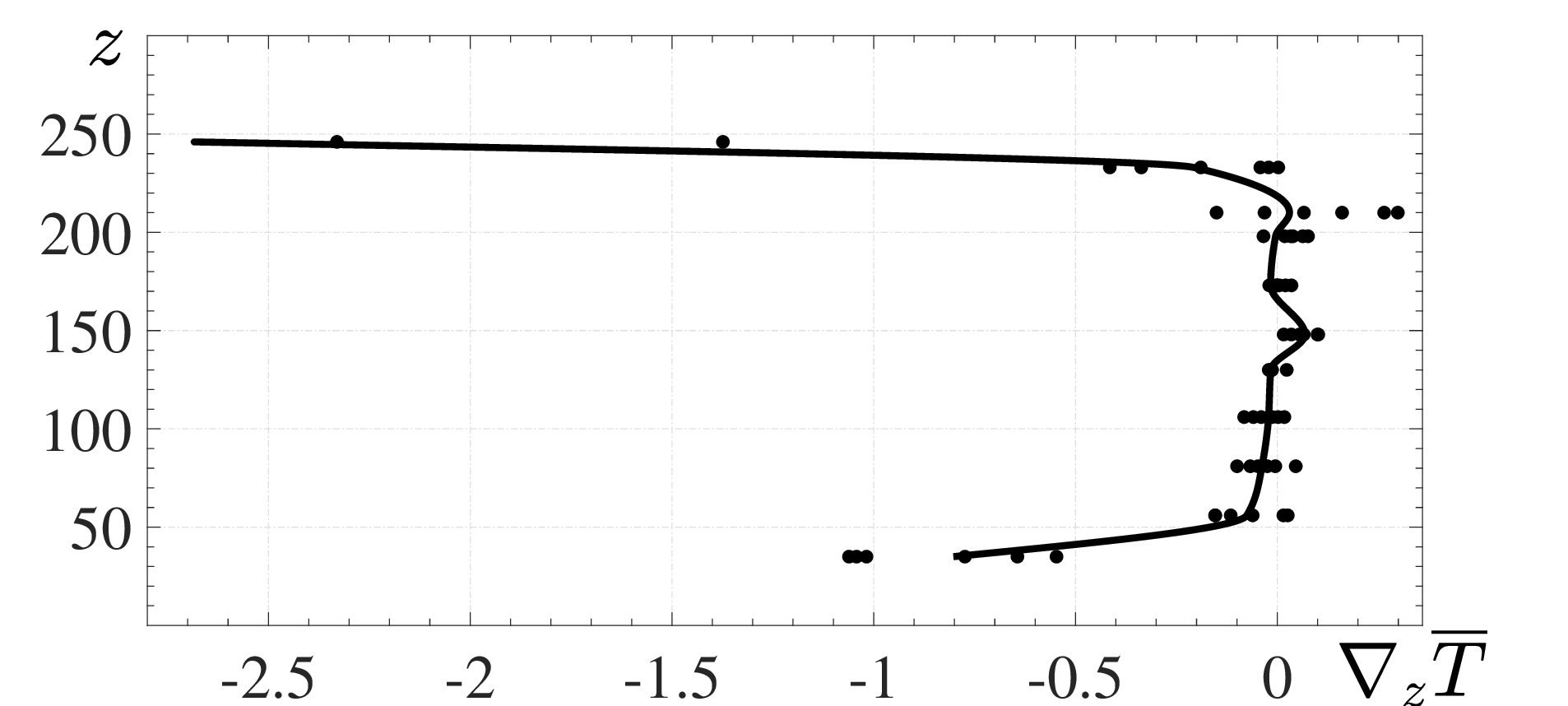}
\caption{\label{Fig9}
Mean profile of the vertical mean temperature gradient $\nabla_z \overline{T}$
in the experiments with the smooth plane bottom surface for $\Delta T = 50$ K
obtained by averaging over 11 vertical profiles measured in different cross-sections
for different $y$
inside the large-scale circulation.
Here the height $z$ is measured in mm, and
the mean temperature gradient is measured in K/cm.
}
\end{figure}

\begin{figure}
\centering
\includegraphics[width=8.7cm]{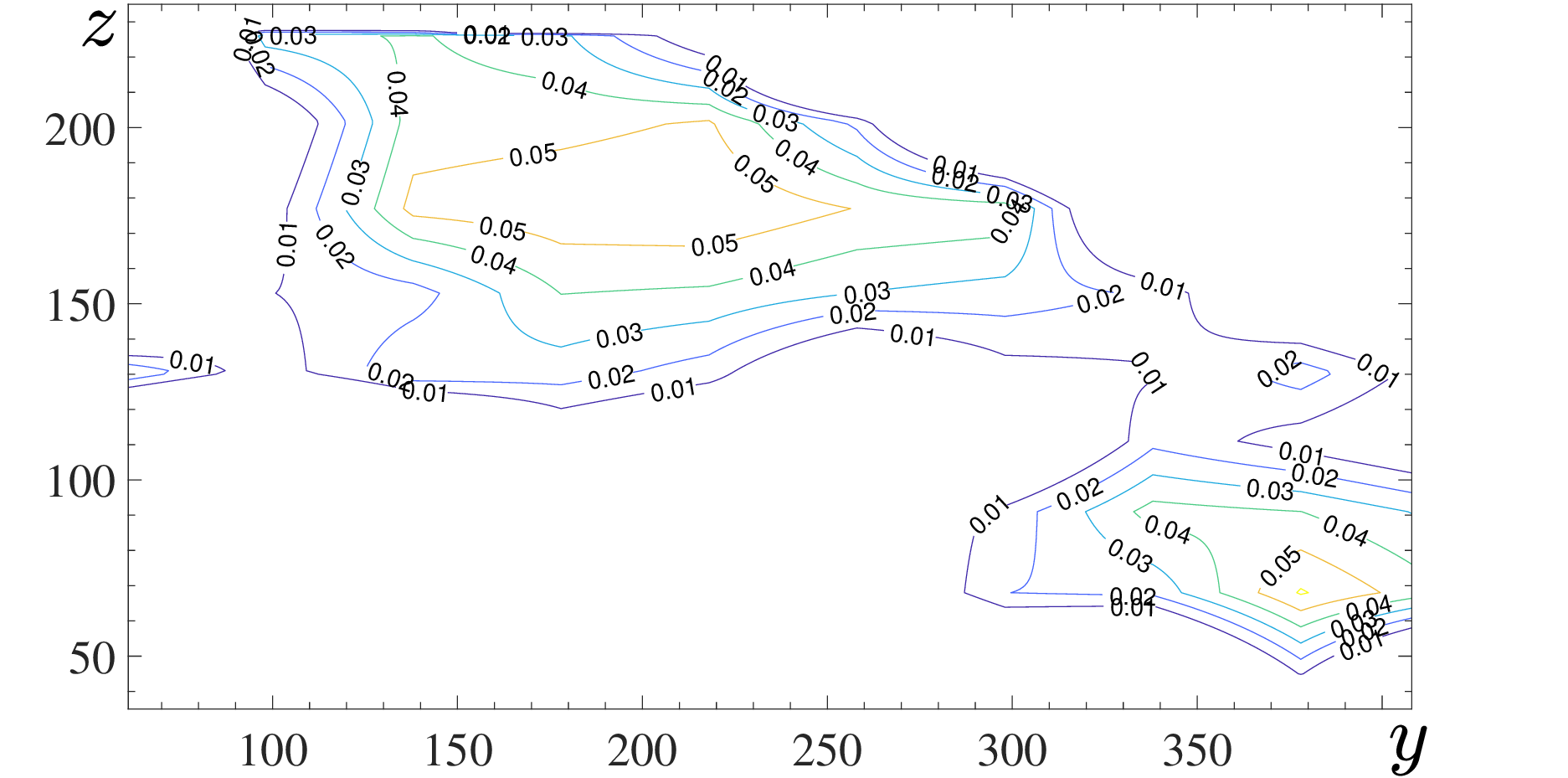}
\caption{\label{Fig10}
Isolines in $yz$ plane of the vertical mean temperature gradient $\nabla_z \overline{T}$
inside the large-scale circulation in the experiments
with the smooth plane bottom surface for $\Delta T = 40$ K.
Only the regions with positive vertical mean temperature gradient are shown.
Here $y$ and $z$ are measured in mm, and
the mean temperature gradient is measured in K/cm.
}
\end{figure}

\begin{figure}
\centering
\includegraphics[width=8.7cm]{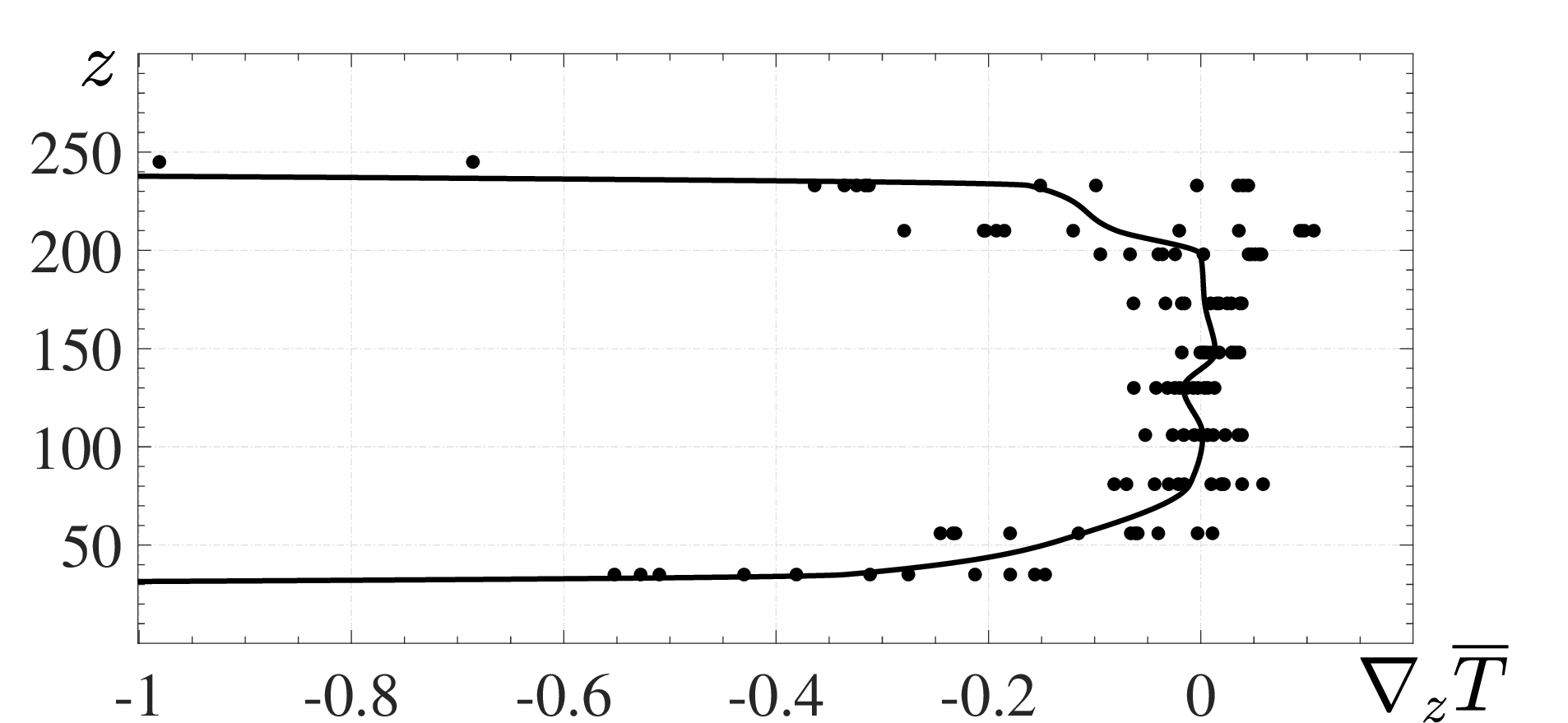}
\caption{\label{Fig11}
Mean profile of the vertical mean temperature gradient $\nabla_z \overline{T}$
in the experiments with the smooth plane bottom surface for $\Delta T = 40$ K
obtained by averaging over 11 vertical profiles measured in different cross-sections
for different $y$
inside the large-scale circulation.
Here the height $z$ is measured in mm, and
the mean temperature gradient is measured in K/cm.
}
\end{figure}

For comparison of results in two types of experiments with wavy and smooth plane bottom surfaces,
in Figs.~\ref{Fig8} and~\ref{Fig10} we show isolines of
the vertical mean temperature gradient $\nabla_z \overline{T}$ in $yz$
plane inside the large-scale circulation obtained in the experiments
with the smooth plane bottom surface.
Similarly, in Figs.~\ref{Fig9} and ~\ref{Fig11} we also show
mean profiles of the vertical mean temperature gradient $\nabla_z \overline{T}$
obtained in these experiments by averaging over 11 vertical profiles
measured in different cross-sections
for different $y$ inside the large-scale circulation.
Here Figs.~\ref{Fig8} and~\ref{Fig9}
correspond to the experiments with $\Delta T = 50$ K, while Figs.~\ref{Fig10} and~\ref{Fig11}
are for the experiments with $\Delta T = 40$ K.
Since the large-scale circulation is not symmetric
due to formation of small additional vortices near the right wall of the chamber
(see Fig.~\ref{Fig3} at $y > 400$ mm),
we show in Figs.~\ref{Fig4}, \ref{Fig6}, \ref{Fig8} and~\ref{Fig10}
the isolines of the vertical mean temperature gradient $\nabla_z \overline{T}$
inside the large-scale circulation only in the range of $y \leq 400$ mm.

Figures~\ref{Fig4}--\ref{Fig11} demonstrate that the spatial distributions of the vertical mean
temperature gradient $\nabla_z \overline{T}$
in the experiments with the smooth plane bottom surface
are slightly different from those obtained in the experiments with the wavy bottom surface.
In particular, in the experiments with the smooth plane bottom surface,
there are less locations with positive vertical mean
temperature gradient $\nabla_z \overline{T}$ than those
in the experiments with the wavy bottom surface.

To study internal gravity waves, we determine spectrum of the long-term variations
of the temperature field characterising the large-scale waves.
From this spectral analysis we obtain main frequencies of internal gravity waves.
The spectrum function $E_{T}(f) = \tilde T^{\rm W}(f) \, \tilde T^{{\rm W}\ast}(f)$ 
for the temperature field (containing 500 frequency data points) has been determined
at 80 locations of the stably stratified region, where
$\tilde T^{\rm W}(f)$ is the Fourier component of the temperature $T^{\rm W}(t)$. 
For every frequency $f$, the obtained spectrum functions $E_{T}(f)$ have been averaged
over 80 locations.

\begin{figure}
\centering
\includegraphics[width=8.5cm]{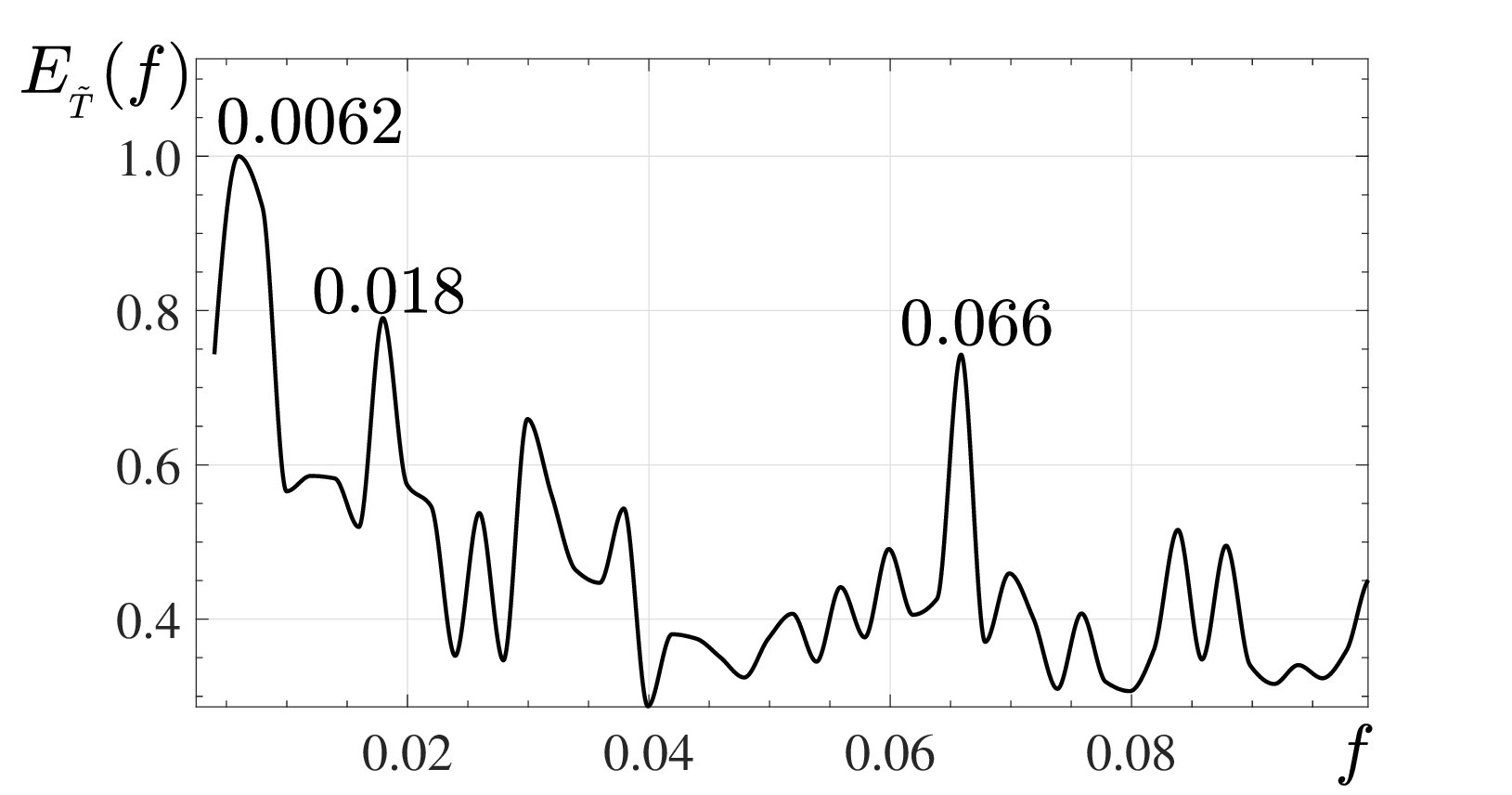}
\caption{\label{Fig12}
The averaged spectrum function $E_{\tilde T}(f)$ of the temperature field
obtained in the experiments with the wavy bottom surface
for the temperature differences $\Delta T= 50$ K.
The main frequencies of the large-scale internal gravity waves
measured in Hz are indicated above the maxima of $\tilde T$.
The function $\tilde T(f)$ is measured in $K^2$/Hz.
}
\end{figure}

\begin{figure}
\centering
\includegraphics[width=9.5cm]{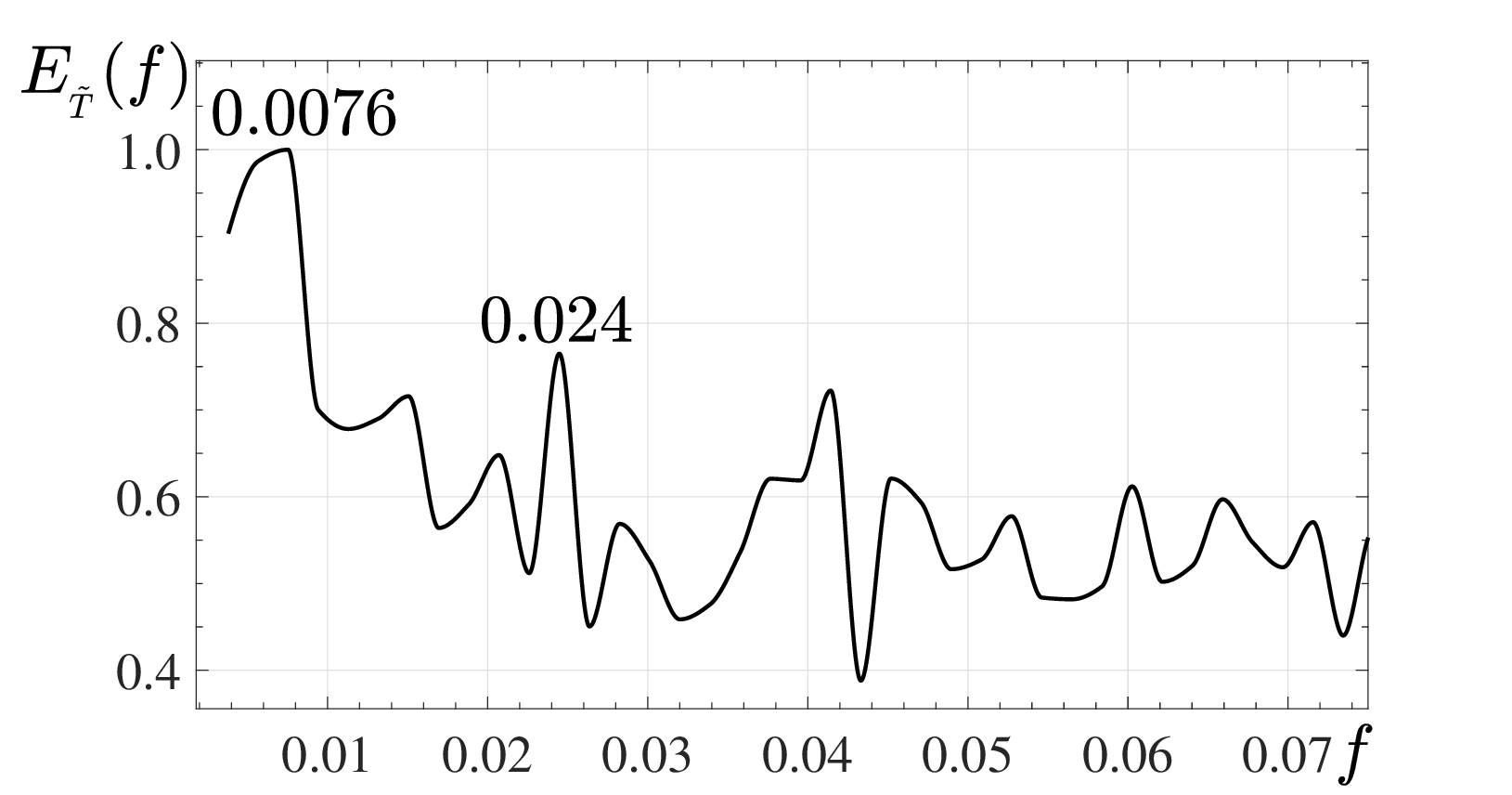}
\caption{\label{Fig13}
The averaged spectrum function $E_{\tilde T}(f)$ of the temperature field
obtained in the experiments with the wavy bottom surface for the temperature differences $\Delta T= 40$ K.
The main frequencies of the large-scale internal gravity waves measured in Hz are
indicated above the maxima of $\tilde T$.
The function $\tilde T(f)$ is measured in $K^2$/Hz.
}
\end{figure}

\begin{figure}
\centering
\includegraphics[width=8.5cm]{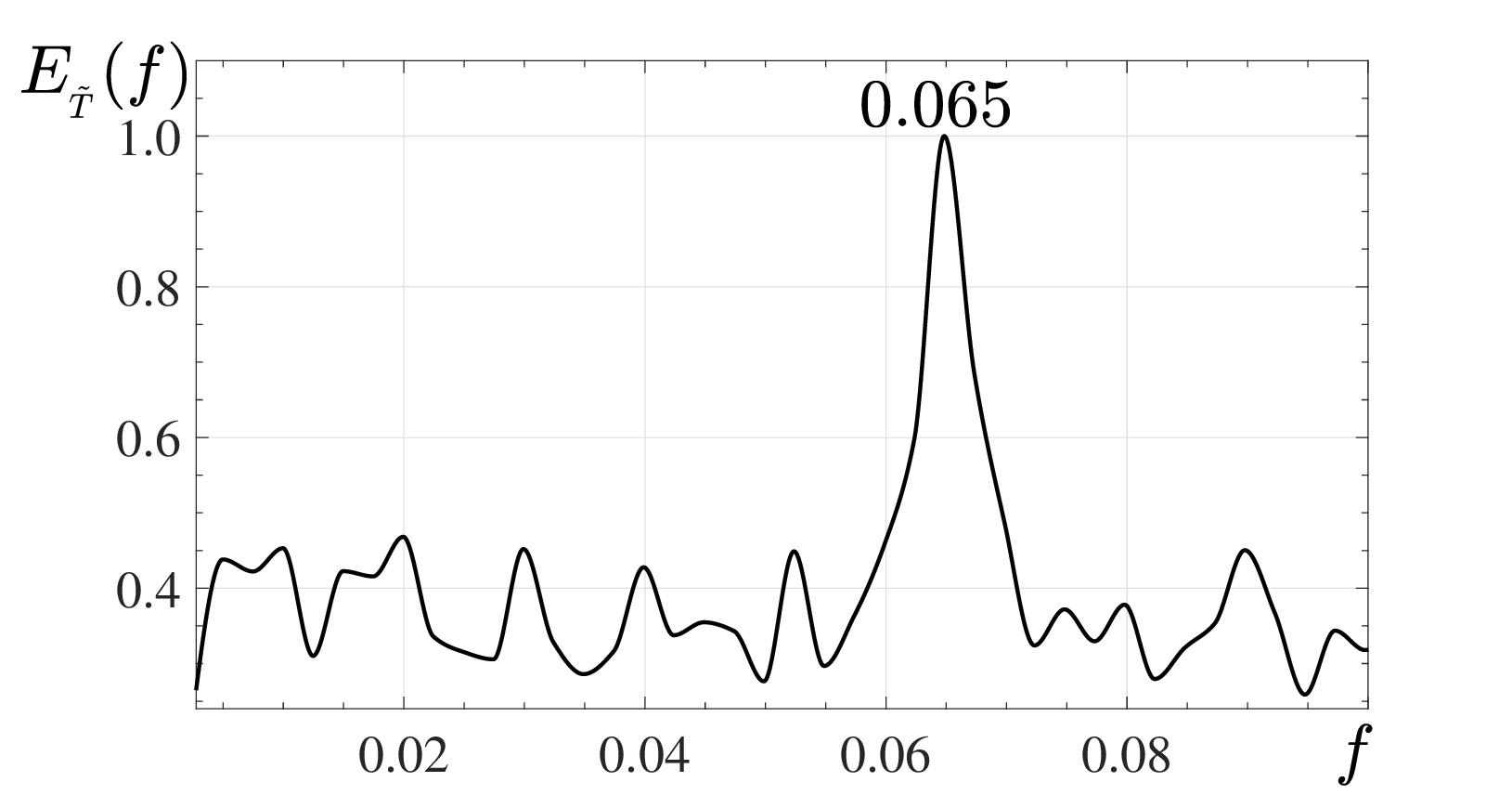}
\caption{\label{Fig14}
The averaged spectrum function $E_{\tilde T}(f)$ of the temperature field
obtained in the experiments with the smooth bottom surface
for the temperature differences $\Delta T= 50$ K.
The main frequencies of the large-scale internal gravity waves
measured in Hz are indicated above the maxima of $\tilde T$.
The function $\tilde T(f)$ is measured in $K^2$/Hz.
}
\end{figure}

\begin{figure}
\centering
\includegraphics[width=8.5cm]{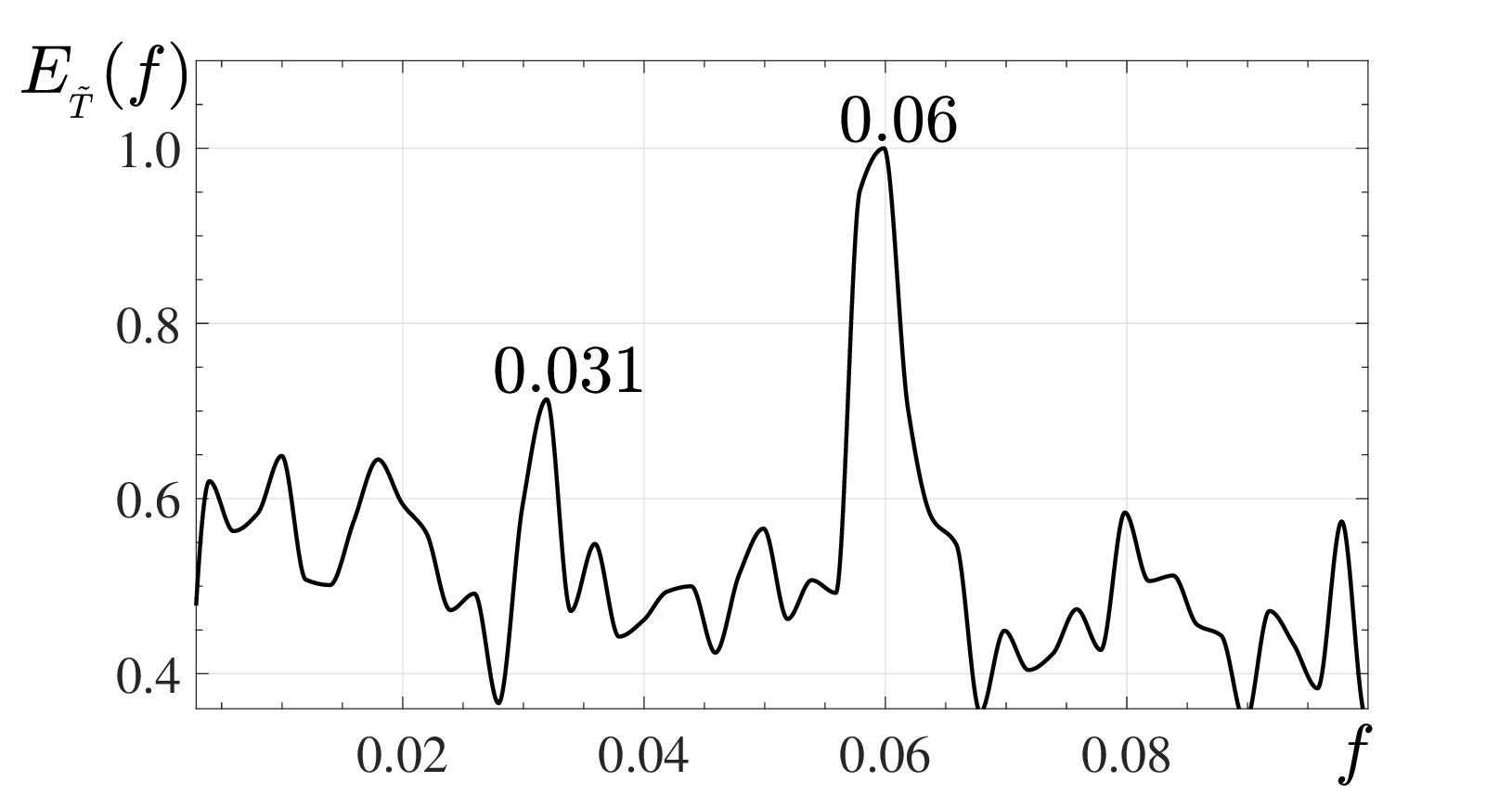}
\caption{\label{Fig15}
The averaged spectrum function $E_{\tilde T}(f)$ of the temperature field
obtained in the experiments with the smooth bottom surface
for the temperature differences $\Delta T= 40$ K.
The main frequencies of the large-scale internal gravity waves
measured in Hz are indicated above the maxima of $\tilde T$.
The function $\tilde T(f)$ is measured in $K^2$/Hz.
}
\end{figure}

In Figs.~\ref{Fig12} and~\ref{Fig13} we show
the volume averaged spectrum function $E_{\tilde T}(f) = \langle E_{T}(f)\rangle_{\rm vol}$
of the temperature field in the experiments with the wavy bottom surface
for the temperature differences $\Delta T= 50$ K (Fig.~\ref{Fig12}) and
$\Delta T= 40$ K (Fig.~\ref{Fig13}),
where the brackets $\langle ... \rangle_{\rm vol}$ denote the averaging
over 80 locations.
For comparison, in Figs.~\ref{Fig14} and~\ref{Fig15} we show
the volume averaged spectrum function $E_{\tilde T}(f) = \langle E_{T}(f)\rangle_{\rm vol}$
of the temperature field in the experiments with the smooth plane bottom surface
for $\Delta T= 50$ K (Fig.~\ref{Fig14}) and
$\Delta T= 40$ K (Fig.~\ref{Fig15}).
It is seen in Figs.~\ref{Fig12}--\ref{Fig15}, that
the measured spectrum contains several localized maxima
in all experiments with the wavy and smooth plane bottom surfaces,
which can be interpreted as large-scale standing internal gravity waves
excited in the regions with the stably stratified flow.

\begin{figure}
\centering
\includegraphics[width=8.5cm]{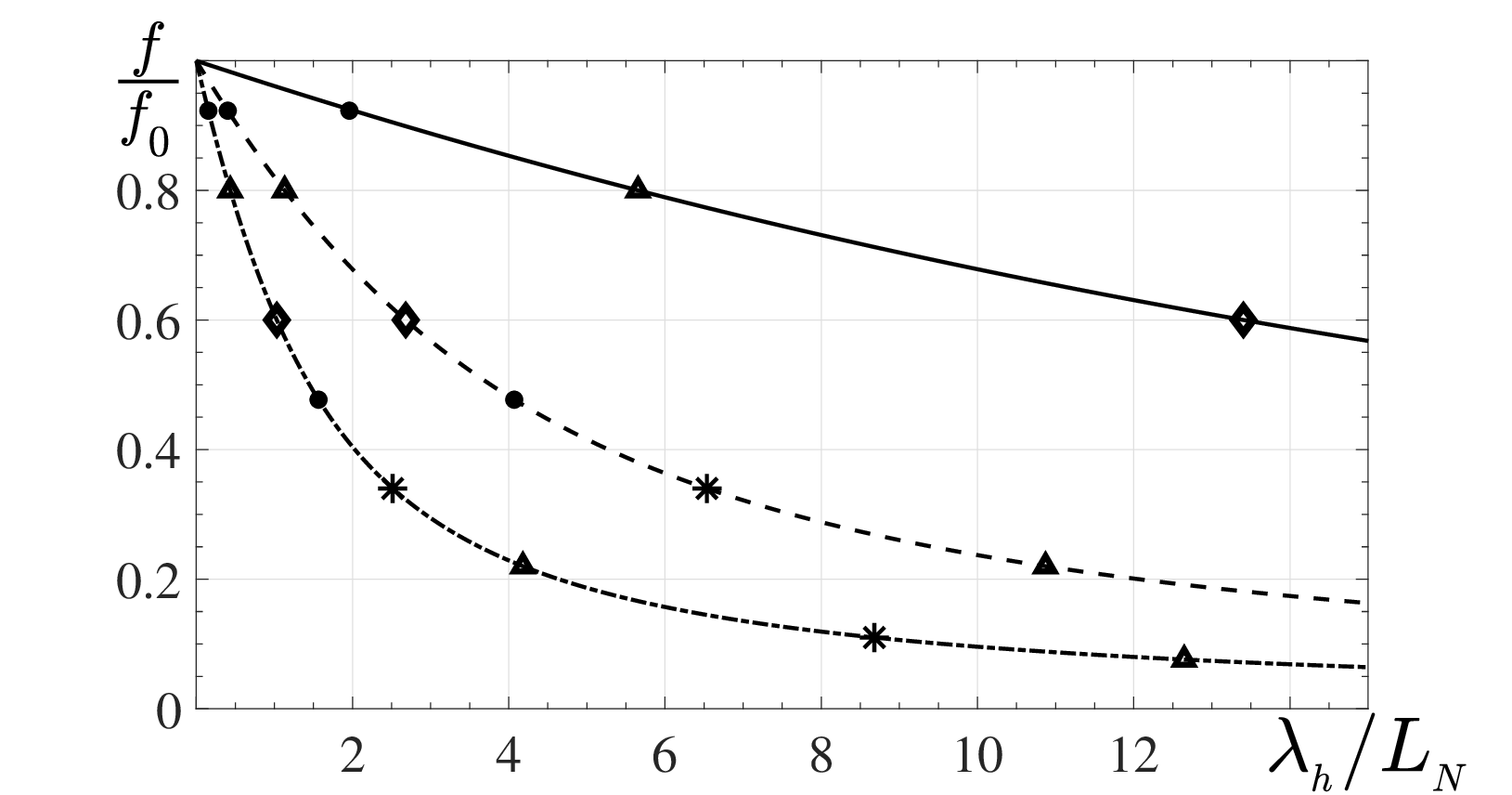}
\caption{\label{Fig16}
The normalized frequencies $f/f_0$ of the internal gravity waves versus the normalized horizontal wavelength $\lambda_h/L_N$: theoretical curves for different modes $m=0$ (solid), $m=1$ (dashed) and $m=3$ (dashed-dotted) and measured frequencies in the experiments with the wavy bottom surface for the temperature differences $\Delta T= 50$ K (triangles) and $\Delta T= 40$ K (snowflakes), and in the experiments with the smooth bottom surface
for $\Delta T= 50$ K (diamonds) and $\Delta T= 40$ K (circles).
}
\end{figure}

Figure~\ref{Fig16} show the theoretical dependencies for the normalised frequencies $f/f_0$ of
the standing internal gravity waves versus the normalized horizontal wavelength
$\lambda_h/L_N$ given by Eq.~(\ref{B14}),
where $L_N \approx 10$ cm is the characteristic vertical scale of the
Brunt-V\"{a}s\"{a}la frequency variations, $\lambda_h=2 \pi/k_h$ is the horizontal wavelength and
$f_0= N_0/2\pi$. The theoretical curves are plotted for the main modes of the standing internal gravity waves.
When $m>2$, the theoretical curves are located very close to each other, so we plot in Fig.~\ref{Fig16} 
the theoretical curves for $m=0, 1, 3$.
Comparing these theoretical curves and the measured frequencies
of the large-scale standing internal gravity waves
obtained in the temperature measurements, we determine the
horizontal wave-numbers for the observed large-scale standing internal gravity waves.

In Fig.~\ref{Fig16} we show the measured frequencies of the large-scale standing internal gravity waves in the experiments with the wavy and smooth plane bottom surfaces for the temperature differences $\Delta T= 50$ K and $\Delta T= 40$ K.
In the experiments with the wavy bottom surface, the frequency $f_0$ based on
the maximum Brunt-V\"{a}s\"{a}la is $f_0=8.2 \times 10^{-2}$ Hz for $\Delta T= 50$ K and $f_0=7.1 \times 10^{-2}$ Hz for $\Delta T= 40$ K. In the experiments with the smooth plane bottom surface, the frequency $f_0=1.08 \times 10^{-1}$ Hz for $\Delta T= 50$ K and $f_0=6.5 \times 10^{-2}$ Hz for $\Delta T= 40$ K.
As follows from Figs.~\ref{Fig12}--\ref{Fig15}, in all experiments there are 8 main frequencies of the large-scale standing internal gravity waves. Since we show in Fig.~\ref{Fig16} only the limited range of the horizontal wavelength, $0 < \lambda_h/L_N < 18$ of the standing internal gravity waves, for higher frequencies, $0.6 < f/f_0 < 1$, we observe three modes for $m=0, 1, 3$, for intermediate range, $0.1 < f/f_0 < 0.5$, we observe two modes for $m= 1, 3$ and for lower frequencies,
$f/f_0 < 0.1$, we observe only one mode for $m= 3$.

The measured wavelength and the period of these waves
are much larger than the turbulent spatial and time scales, respectively.
In particular, the frequencies of the observed large-scale waves vary
from 0.006 Hz to 0.07 Hz in the experiments with the wavy bottom surface
and from 0.03 Hz to 0.065 Hz in the experiments with the smooth plane bottom surface,
while the turbulent time in the integral scale
is about 0.5 s.
Note that inside the chamber, we observe standing but not propagating internal gravity waves
because the chamber is a small closed volume.
Actually, interactions of propagating and reflecting waves
cause the standing internal gravity waves,
which are observed in our experiments.

\begin{figure}
\centering
\includegraphics[width=7.5cm]{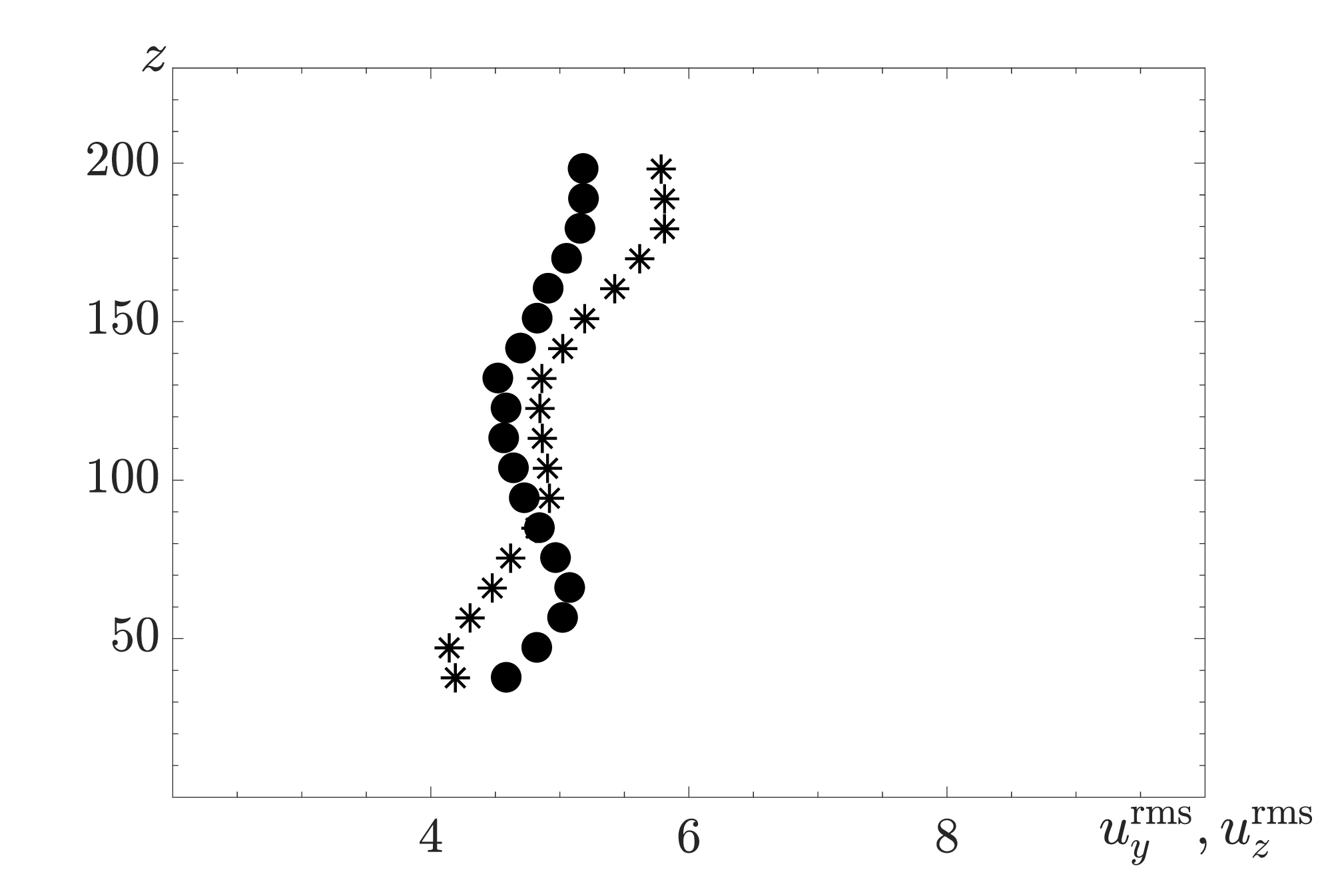}
\includegraphics[width=7.5cm]{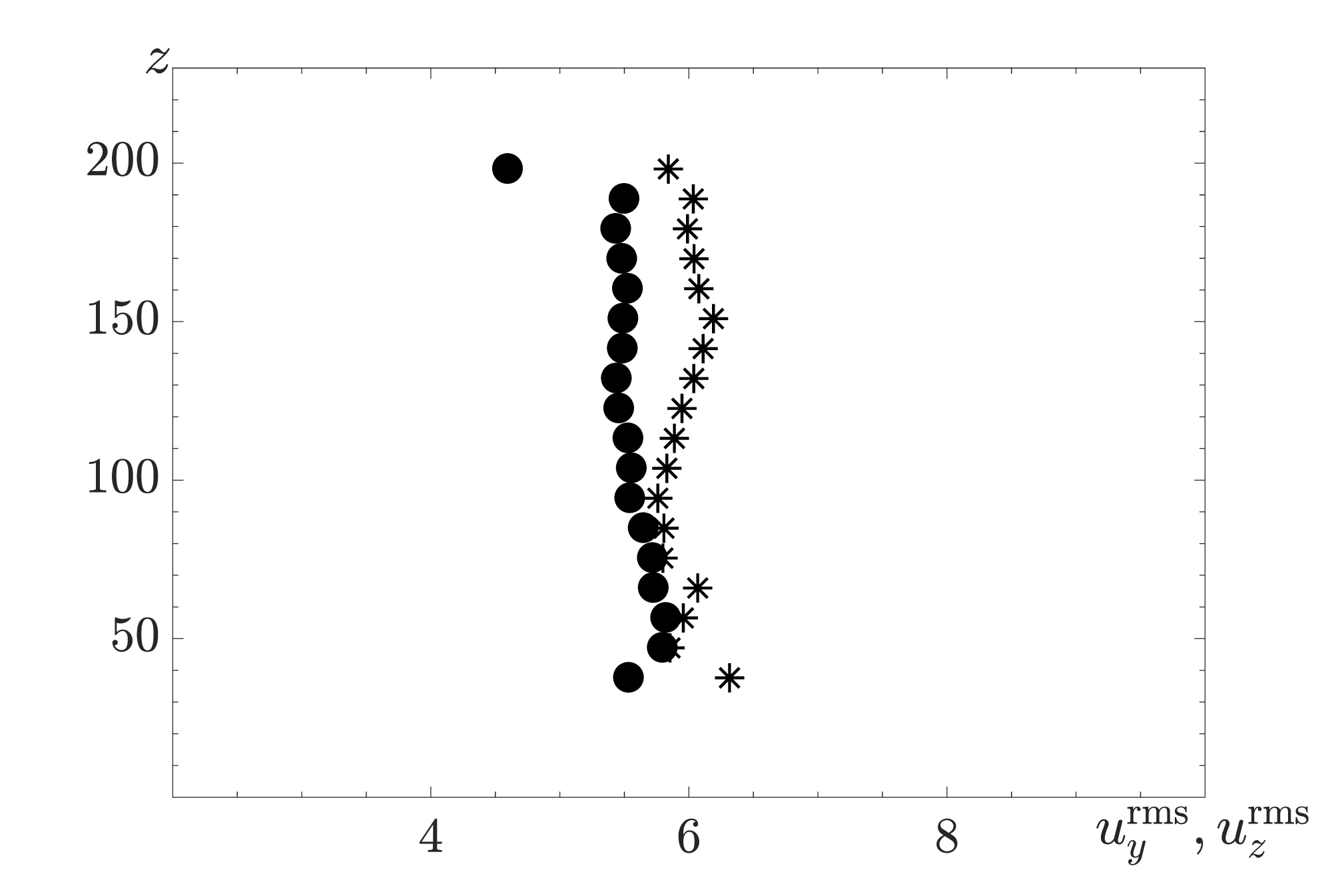}
\caption{\label{Fig17}
The vertical profiles of the rms velocities $u_y^{\rm rms}$ (snowflakes)
and $u_z^{\rm rms}$ (circles) in the experiments with the wavy bottom surface
(upper panel) and the smooth plane bottom surface (bottom panel)
obtained at $\Delta T= 50$ K.
Here the $z$ is measured in mm, and
the rms velocities $u_y^{\rm rms}$
and $u_z^{\rm rms}$ are  measured in cm/s.
}
\end{figure}

We also compare the results of measurements of velocity fluctuations 
in the two sets of experiments
with the wavy and smooth plane bottom surfaces,
conducted at the same mean temperature difference
between bottom and upper surfaces. In Fig.~\ref{Fig17} we show 
the vertical profiles of the rms velocity components
$u_y^{\rm rms}=\sqrt{\langle u_y^2 \rangle}$ and
$u_z^{\rm rms}=\sqrt{\langle u_z^2 \rangle}$,
where the spatial average is over the reduced domain 
(excluding the end boundaries).
In the experiments with the wavy bottom surface,
there is a minimum in the vertical profile of the rms velocities
in the center part of the chamber.
On the other hand, in the experiments with the smooth plane bottom surface,
the vertical distribution of the rms velocities is essentially
less inhomogeneous.
Also the values of turbulent velocities in the experiments
with the wavy bottom surface are smaller compared to velocity fluctuations in
the experiments with the smooth plane bottom surface.

The reason for these differences may be as follows.
In the experiments with the wavy bottom surface,
the intensity of the velocity fluctuations
are partially depleted by the production of the temperature fluctuations
in stably stratified regions, where the part of turbulent kinetic energy
is transformed to the turbulent potential energy caused by the
temperature fluctuations. \cite{ZKR07,ZKR13}
In another words, in stably stratified flows, the production of turbulent kinetic energy
is depleted by destruction caused by the buoyancy. 
On the other hand, in the experiments with the smooth plane bottom surface,
there are less locations of the stably stratified regions, and
the level of velocity fluctuations is larger in comparison with that
observed in the experiments with the wavy bottom surface.

\section{Conclusions}

In the present study we perform laboratory experiments in
convection over a wavy heated bottom surface.
An interaction of the large-scale circulation
with the wavy heated bottom surface strongly affects
the spatial structure of the mean temperature field
and generation of the large-scale standing internal
gravity waves in the flow core. In particular,
we have found that there are many locations with stably stratified
regions in the flow core of the large-scale circulation,
and the large-scale standing internal
gravity waves are observed in these regions.
The wavelength and the period of these waves
are much larger than the turbulent spatial and time scales.
The spectrum of these waves contains several localized maxima, that
is an indication of existence of the waveguide resonators for the internal
gravity waves.

In the experiments with a smooth plane bottom surface
at the same temperature difference
between bottom and upper surfaces, there are 
less locations with a stably stratified turbulence.
However, the large-scale standing internal gravity waves 
are detected in both experiments
with the wavy and smooth plane bottom surfaces.

The turbulence in the region with stably stratified flows in the experiments
with the wavy bottom surface is inhomogeneous,
e.g., there is a minimum in the vertical profiles of the rms velocities
in the center part of the chamber, where the intensity of the velocity fluctuations
are partially depleted by the production of the temperature fluctuations.
On the other hand, in the experiments with the smooth plane bottom surface,
the vertical distribution of the rms velocities is nearly homogeneous.

\begin{acknowledgements}
This paper is dedicated to Prof. T. Elperin (1949-2018) who initiated this work.
We thank A.~Krein for his assistance in construction of the
experimental set-up and E.~Elmakies for his assistance in data analysis.
The detailed comments on our manuscript by the anonymous referees 
are very much appreciated. 
This research was supported in part by the Israel Ministry of Science and Technology (grant No. 3-16516)
and the PAZY Foundation of the Israel Atomic Energy Commission (IAEC) (Grant No. 122-2020).
\end{acknowledgements}

\bigskip
\noindent
{\bf DATA AVAILABILITY}
\medskip

The data that support the findings of this study are available from the corresponding author
upon reasonable request.

\appendix
\section{Nomenclature}

$c_{\rm p}$ is the specific heat at constant pressure,

$c_{\rm v}$ is the specific heat at constant volume,

${\bm e}$ is the vertical unit vector,

$E_{T}(f)$ is the spectrum function for the temperature field,

$E_{\tilde T}(f)$ is the volume averaged spectrum function for the temperature field,

$f$ is the frequency,

${\bm g}$ is the acceleration due to gravity,

${\bm k}$ is the wave vector,

${\bm k}_h=(k_x, k_y)$ is the horizontal component of the wave vector,

$k_z$ is the vertical component of the wave vector,

$\ell$ is the integral scale of turbulence,

$L_N$ is the characteristic vertical scale of the Brunt-V\"{a}is\"{a}l\"{a} frequency variations,

$L_x$ and $L_y$ are the horizontal sizes of the chamber along the $x$ and $y$ axes,

$L_z$ is the height measured from the lower point of a wavy surface
with the sinusoidal modulation to the upper surface of the chamber,

$N(z)= (g \nabla_z S_{\rm eq})^{1/2}$ is Brunt-V\"{a}is\"{a}l\"{a} frequency,

$P_{\rm eq}$ is the pressure at an equilibrium,

$P^{\rm W}$ are the pressure wave field
(the long-term variations of the pressure
due to the large-scale internal gravity
waves),

${\bm r}$ is the radius-vector of the centre of the wave packet,

${\rm Ra}$ is the Rayleigh number,

${\rm Ri}$ is the gradient Richardson number,

$S_{\rm eq}$ is the entropy at an equilibrium,

$S^{\rm W}$ are the entropy wave field
(the long-term variations of the entropy
due to the large-scale internal gravity
waves),

${\rm Sh}$ is the mean velocity shear of the large-scale circulation,

$T^{\rm tot}=\overline{T} + T^{\rm W} + \theta$ is the total instantaneous temperature,

$T=\overline{T} + T^{\rm W}$ is the temperature determined by sliding averaging of the
instantaneous total temperature field over the time (5 s),
that is by one order of magnitude larger than the
characteristic turbulence time (0.5 s),

$\overline{T}$ is the mean temperature obtained by the
averaging of the temperature $T$ over the time 500~s,

$T_{\rm eq}$ is the temperature at an equilibrium,

$T^{\rm W}=T-\overline{T}$ is the temperature wave field
(the long-term variations of the temperature
due to the large-scale internal gravity
waves),

$\tilde T^{\rm W}(f)$ is the Fourier component 
of the temperature $T^{\rm W}(t)$,

$\Delta T$ is the temperature difference
between the bottom and upper surfaces of the chamber,

$V_\ast$ is the amplitude of the vertical velocity for the wave field,

${\bm V}_{\rm eq}$ is the velocity at an equilibrium,

${\bm V}^{\rm W}$ are the velocity wave field
(the long-term variations of the velocity
due to the large-scale internal gravity
waves),

${\bm V}^{\rm W}_\perp=(V^{\rm W}_x, V^{\rm W}_y)$ is the horizontal component of the velocity wave field,

${\bf u}$ are the velocity fluctuations,

$u_y^{\rm rms}$ is the rms of the turbulent horizontal velocity,

$u_z^{\rm rms}$ is the rms of the turbulent vertical velocity,

$z_{\rm max}$ is the reflection (or ''turning") point,

$\alpha$ is the thermal expansion coefficient,

$\gamma=c_{\rm p}/c_{\rm v}$ is the ratio of the specific heats,

$\theta$ are the temperature fluctuations,

$\kappa$ is the thermal diffusivity,

$\lambda_h$ is the horizontal wavelength,

$\nu$ is the kinematic viscosity,

$\rho_{\rm eq}$ is the density at an equilibrium,

$\tau$ is the turbulence time scale at the integral scale of turbulence,

$\varphi$ is the phase of the velocity  perturbations,

$\omega$ is the frequency of the internal gravity waves.

\end{document}